\documentclass[preprint2]{aastex631}

\usepackage[utf8]{inputenc}
\usepackage{gensymb}

\usepackage{graphicx}
\usepackage{natbib}
\usepackage{amsmath}

\newcommand{\sco}{S$_{\mathrm{CO}} \Delta \mathrm{v}$}
\newcommand{\iras}{IRAS F01364$-$1042}
\newcommand{\zw}{III Zw 035}
\newcommand{\msun}{M$_\odot$}
\newcommand{\lsun}{L$_\odot$}
\newcommand{\menc}{M$_{\mathrm{enc}}$}

\newcommand{\mgasco}{M$_{\mathrm{gas, CO}}$}
\newcommand{\mgasconthundred}{M$_{\mathrm{gas, cont (100 K)}}$}
\newcommand{\mgascontfhundred}{M$_{\mathrm{gas, cont (500 K)}}$}
\newcommand{\mbh}{M$_\mathrm{BH}$}

\begin{document}

\title{Constraining Nuclear Molecular Gas Content with High-resolution CO Imaging of GOALS Galaxies}
\author[0000-0002-8122-3032]{James Agostino}
\affiliation{Ritter Astrophysical Research Center and Department of Physics \& Astronomy, University of Toledo, Toledo, OH 43606, USA}
\author[0000-0001-7421-2944]{Anne M. Medling}
\affiliation{Ritter Astrophysical Research Center and Department of Physics \& Astronomy, University of Toledo, Toledo, OH 43606, USA}
\author[0000-0003-0057-8892]{Loreto Barcos-Mu\~{n}oz} 
\affiliation{National Radio Astronomy Observatory, 520 Edgemont Road, Charlottesville, VA, 22903, USA}
\affiliation{Department of Astronomy, University of Virginia, 530 McCormick Road, Charlottesville, VA, 22903, USA}
\author[0000-0002-1912-0024]{Vivian U}
\affiliation{4129 Frederick Reines Hall, Department of Physics and Astronomy, University of California, Irvine, CA 92697, USA}
\author{Mynor Rodr\'{i}guez V\'{a}squez}
\affiliation{Instituto de Investigaci\'{o}n en Ciencias F\'{i}sicas y Matem\'{a}ticas USAC, Ciudad Universitaria, Zona 12, 01012, Guatemala, Guatemala}
\author[0000-0003-3474-1125]{George C. Privon}
\affiliation{National Radio Astronomy Observatory, 520 Edgemont Road, Charlottesville, VA 22903}
\affiliation{Department of Astronomy, University of Florida, P.O. Box 112055, Gainesville, FL 32611, USA}
\affiliation{Department of Astronomy, University of Virginia, 530 McCormick Road, Charlottesville, VA 22904, USA}
\author[0000-0003-0522-6941]{Claudia Cicone}
\affiliation{Institute of Theoretical Astrophysics, University of Oslo, PO Box 1029, Blindern, 0315 Oslo, Norway}
\author[0000-0003-3498-2973]{Lee Armus}
\affiliation{IPAC, California Institute of Technology, 1200 East California Boulevard, Pasadena, CA 91125, USA}
\author[0000-0002-3430-3232]{Jorge Moreno}
\affiliation{Department of Physics and Astronomy, Pomona College, Claremont, CA 91711, USA}
\affiliation{The Observatories of the Carnegie Institution for Science, 813 Santa Barbara Street, Pasadena, CA 91101, USA}
\author[0000-0001-5231-2645]{Claudio Ricci}
\affiliation{Instituto de Estudios Astrof\'{\i}sicos, Facultad de Ingenier\'{\i}a y Ciencias, Universidad Diego Portales, Avenida Ejercito Libertador 441, Santiago, Chile}
\affiliation{Kavli Institute for Astronomy and Astrophysics, Peking University, Beijing 100871, China}
\author[0000-0002-3139-3041]{Yiqing Song}
\affiliation{European Southern Observatory, Alonso de Córdova, 3107, Vitacura, Santiago, 763-0355, Chile}
\affiliation{Joint ALMA Observatory, Alonso de Córdova, 3107, Vitacura, Santiago, 763-0355, Chile}
\author[0000-0003-4073-3236]{Christopher C. Hayward}
\affiliation{Eureka Scientific, Inc., 2452 Delmer Street, Suite 100, Oakland, CA 94602, USA}
\affiliation{Kavli Institute for the Physics and Mathematics of the Universe (WPI), The University of Tokyo Institutes for Advanced Study, The University of Tokyo, Kashiwa, Chiba 277-8583, Japan}
\affiliation{Center for Computational Astrophysics, Flatiron Institute, 162 Fifth Avenue, New York, NY 10010, USA}
\author[0000-0002-4261-2326]{Katherine Alatalo}
\affiliation{Space Telescope Science Institute, 3700 San Martin Drive, Baltimore, MD 21211, USA}
\affiliation{William H. Miller III Department of Physics and Astronomy, Johns Hopkins University, Baltimore, MD 21218, USA}
\author[0000-0002-1233-9998]{David B. Sanders}
\affiliation{Institute for Astronomy, University of Hawaii, 2680 Woodlawn Drive, Honolulu, HI 96822}
\onecolumngrid

\begin{abstract}

We present measurements of the cool molecular gas mass around the nuclei of two gas-rich mergers, \zw\ and \iras, whose enclosed masses (M$_\mathrm{enc}$) within the central 40-80 pc would be overmassive if attributed entirely to the supermassive black hole mass (SMBH) and compared to SMBH-galaxy scaling relations. Our gas mass measurements are derived from Atacama Large Millimeter/submillimeter Array (ALMA) Band 6 long-baseline observations of CO(J=2-1) and 230 GHz continuum emission at 14-20 pc resolution, which probes below the resolving limit of the previous black hole mass measurements. Subtracting molecular gas mass from these enclosed masses is not enough to reconcile with BH-galaxy relationships, but independently measuring \menc\  using the cold CO(2-1) gas does shift the black holes down to their expected values. Still, these ALMA data reveal respective molecular gas masses of $\sim$3$\times$10$^7$ to $\sim$6$\times$10$^8$ \msun\ within 70 pc of these black holes, which could challenge some black hole accretion models that assume nuclear gas like this has no angular momentum.

\end{abstract}

\section{Introduction}

Every massive galaxy in our universe is thought to host a supermassive black hole (SMBH) in its center. These SMBHs typically contain masses millions or billions times the mass of our Sun. Their gravitational sphere of influence is tiny (of order $\sim$1-100 pc, the radius of which can be defined as r$_{\mathrm{SOI}}$ $\approx$ G\mbh/$\sigma^2_*$; see \citealt{Yoon2017}) when compared to the typical kpc to hundreds of kpc scale sizes of galaxies. Despite this, a number of galaxy-wide properties correlate with the mass of the central SMBH. These are the so-called black hole mass (\mbh) scaling relations, where SMBH mass scales with host galaxy properties like bulge luminosity (\citealt{Kormendy1993a}; \citealt{Ho1999a}; \citealt{MarconiHunt2003}; \citealt{McConnellMa2013}), stellar mass (\citealt{SilkRees1998}; \citealt{Magorrian1998}; \citealt{KormendyGebhart2001}; \citealt{MclureDunlop2002}; \citealt{HaringRix2004}; \citealt{Bennert2011}; \citealt{Cisternas2011a}), and bulge velocity dispersion (\citealt{Gebhardt2000}; \citealt{MerrittFerrarese2001}; \citealt{Tremaine2002}; \citealt{Gultekin2009c}; \citealt{Beifioiri2012}; \citealt{McConnellMa2013}). The mechanism driving these relations is not well understood and the search for a physically-motivated picture for the co-evolution is an active area of research (see \citealt{Kormendy2013} and \citealt{Heckman2014} for reviews and \citealt{Bennert2021} for a recent study on local galaxies). 

One of the major uncertainties in placing galaxies on the SMBH-galaxy scaling relations is a precise measurement of the \mbh. Even with very-long-baseline interferometry, the region near the event horizon can only be directly ``observed'' around the nearest, most massive SMBHs (\citealt{EHT2019}). Depending on the target and data available, multiple approaches can be taken to measure SMBH mass.  Some of these methods include reverberation mapping (\citealt{1991ApJ...366...64C}; \citealt{2004PASP..116..465H}; \citealt{2006ApJ...641..689V}; \citealt{2021iSci...24j2557C}), Schwarzschild modeling (\citealt{Schwarzschild1979}; \citealt{Faber1976}; \citealt{Lauer1995}; \citealt{1997ApJ...482L.139K}; \citealt{1999ApJS..124..383C}; \citealt{Gebhardt2000}; \citealt{vandenbosch2008}; \citealt{Walsh2012}), direct measurements of stellar orbits (\citealt{Ghez2008}; \citealt{Genzel2010}), dynamical modeling of stars and gas in circumnuclear disks ($>$1 pc) (\citealt{2008MNRAS.390...71C}; \citealt{Barth2009}; \citealt{Medling2011}; \citealt{U2013}; \citealt{Medling2014}), and empirically-calibrated proxies (e.g. \citealt{vestergaard2006}; \citealt{Morgan2010}; \citealt{Bentz2013}; \citealt{U2022a}).

\par
In this paper, we refer to a parent sample of nine SMBH masses within gas rich mergers derived via gas and stellar kinematics in \cite{Medling2015}. These enclosed mass (\menc, mass under r = 77 mas from the SMBH) measurements use data from Keck/OSIRIS adaptive optics (AO) and target galaxies from the Great Observatory All-sky LIRG Survey (GOALS, \citealt{2009PASP..121..559A}). In the K-band, these data use the CO (2-0) and (3-1) bandheads to trace kinematics of young stars along with emission from Br$\gamma$ and H$_2$.  LIRGs (luminous infrared galaxies) emit an excess of light in the infrared \citep[L$_{IR}$ $>$ 10$^{11}$ L$_\odot$ by definition;][]{SandersMirabel1996} powered by vigorous activity from either star formation or AGN, or both. Based on the dynamical (\menc) measurements in \cite{Medling2015}, individual galaxies like \zw\ and \iras\ are significantly offset (by between $\sim$0.5 - 2.5 dex) from the canonical black hole mass scaling relations: supermassive black hole mass (\mbh) vs. stellar velocity dispersion ($\sigma _\star$), total stellar mass (M$_{\star}$), and bulge luminosity (L$_{\mathrm{H, bulge}}$). Non-LIRG spiral galaxies do not typically lie above scaling relations (\citealt{Davis2018}; \citealt{Davis2019}), and whether most LIRGs do remains in question. Overmassive black holes present a challenge to the canonical understanding of gas-rich mergers, which predicts that these black holes will grow more in an upcoming quasar phase --- placing them in even greater conflict with scaling relations (\citealt{Mirabel1988}; \citealt{Hopkins2008a}; \citealt{Hopkins2008b}).
\par
The results from \cite{Medling2015} were surprising but came with a caveat: the resolution of the Keck/OSIRIS+AO data does not probe all the way down to the expected SMBH r$_{\mathrm{SOI}} \approx$ 10 pc (or less; value estimated for \iras\ based on $\sigma_\star$ found in \citealt{Medling2015} and equation 3 of \citealt{Kormendy2013}). Those dynamically-derived SMBH mass measurements in \cite{Medling2015} are a \menc\ within the highest achievable resolution by OSIRIS at a 35 mas plate scale, 77 mas (a Nyquist sampling rate of 2.2; see \citealt{Medling2019}), meaning they could include significant extended stellar and gaseous mass. This paper is focused on quantifying exactly how much molecular gas mass exists within that limit, then completing an independent \menc\ measurement for comparison. \cite{Medling2019} performed a pilot study focused on measuring molecular gas mass in one of these galaxies that host double nuclei, NGC 6240. Using Atacama Millimeter/submillimeter Array (ALMA) Band 6 CO(J=2-1) and continuum data to measure molecular gas mass, they found that in the northern nucleus, the black hole shifts down to scaling relations after the gas mass correction, as up to 89\% of the previously measured enclosed mass could be attributed to gas rather than the black hole. In contrast, the dynamical mass of the southern nucleus consists of up to only 11\% gas mass, leaving it above scaling relations. The following work estimates the amount of molecular gas mass in two other gas-rich, merging LIRGs with the purpose of understanding whether or not these galaxies host overmassive black holes. Such confirmation would be further evidence for an accretion paradigm in which the SMBH grows before the global galaxy properties in scaling relations, which is contrary to the timeline of accretion in typical systems (\citealt{Hopkins2012}; \citealt{Cen2012}; \citealt{Angles2017}). 

In this Paper we determine whether the amount of gas present within the resolving limit of the original dynamical mass measurements accounts for the observed disagreement with scaling relations. We then measure an independent \menc\ using the CO(2-1) kinematics. This paper is structured as follows. Section~\ref{sec:MeasurementsandImaging} details the CO(2-1) measurements obtained from the ALMA and how the imaging process of these measurements was conducted. Section~\ref{sec:Analysis} explains how we calculated molecular gas mass estimates via both CO(2-1) and ALMA Band 6 continuum. Section~\ref{sec:PVdiagrams} describes independent \menc\ measurements using the same CO(2-1) cubes and how they compare to the \cite{Medling2015} \menc. Section~\ref{sec:Discussion} contextualizes these results within the picture of scaling relations and what those results mean for our current picture of the co-evolution of merging gas-rich galaxies and their SMBHs. 

The two galaxy merger systems featured in this paper are \zw\ and \iras. \zw\ is at $z$ = 0.02744 and R.A. = 01h44m30.50s, dec. =  +17d06m05.0s with a total infrared luminosity $\log (L_{\mathrm{IR}}/L_\odot)$ = 11.64. \iras\ is at z = 0.04823 and R.A. = 01h38m52.921s, dec. =  -10d27m11.42s with a total infrared luminosity $\log (L_{IR}/L_\odot)$ = 11.85 (\citealt{2009PASP..121..559A}). If \zw\ or \iras\ host AGNs, they must be dust-obscured. 

In this work, we use cosmological parameters of H$_0$ = 70 km~s$^{-1}$ Mpc$^{-1}$, $\Omega$$_m$ = 0.28, and $\Omega$$_\Lambda$ = 0.72 (\citealt{2009ApJS..180..225H}). To calculate spatial scales and luminosity distances for \zw \ and \iras \ we use Ned Wright's Cosmology Calculator (\citealt{Wright2006}). \zw\ has a scale of 554 pc arcsec$^{-1}$ and \iras\ has a scale of 1012 pc arcsec$^{-1}$.

\section{Measurements and Imaging}
\label{sec:MeasurementsandImaging}
\begin{figure*}[t]  
    \centering
    \hspace{-0.7cm}
    \includegraphics[width=8.9cm]{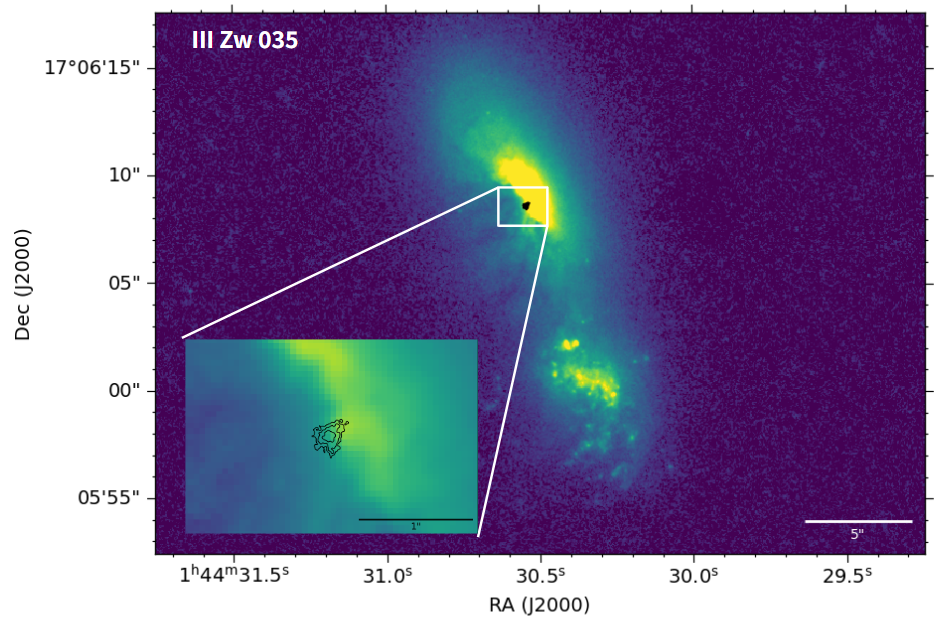}
    \includegraphics[width=8.9cm]{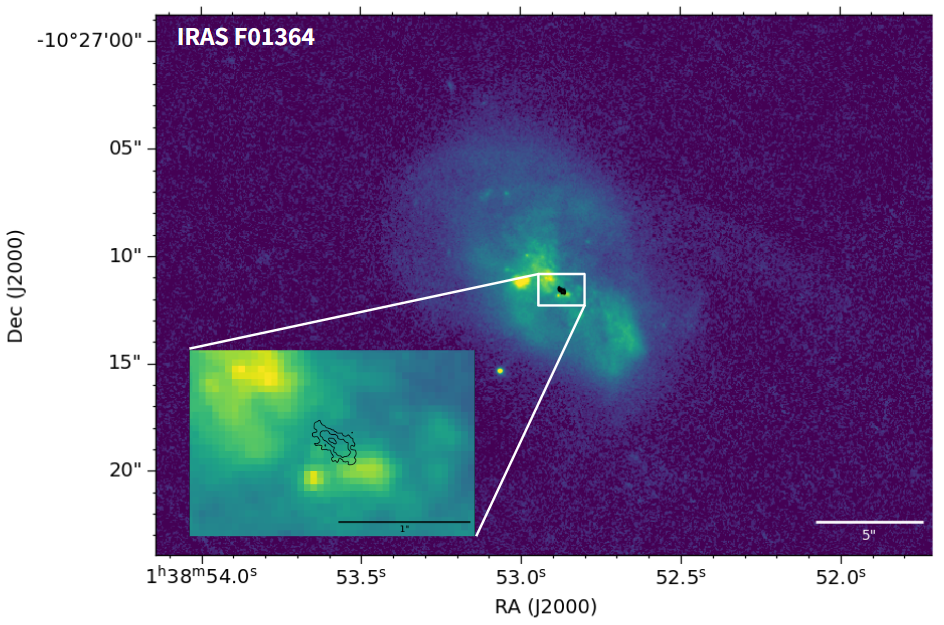}\par\medskip
    \caption{I-Band Hubble ACS images (GO-10592; PI: Evans) of \zw\ (left) and \iras\ (right) with CO(2-1) moment 0 contours overlaid in black. Moment 0 contours are from 10$^{-0.4}$ to 10$^{0.5}$ Jy beam$^{-1}$ km s$^{-1}$ for \zw \ and 10$^{-0.8}$ to 10$^{0.5}$ Jy beam$^{-1}$ km s$^{-1}$ for \iras\ in 7 logarithmically-spaced intervals. \zw's companion galaxy still retains much of its large-scale structure, while in \iras\ the two are in a later stage of the merger process.}
    \label{fig:mosaics}
\end{figure*}

The nuclei of \zw\ and \iras\ were observed with Keck-OSIRIS+AO in 2010 and 2011 (\citealt{Medling2014}). This integral field spectroscopy was used to measure dynamical masses at a resolution of 77 mas (\citealt{Medling2015}). As our science goal is to resolve the inner-most molecular gas in these galactic nuclei, our Band 6 observations were taken in ALMA's second-most extended configuration, C43-9 (PI Medling, project codes 2018.1.01123.S and 2019.1.00811.S). In these programs, there is also $\sim$0.14\arcsec\ resolution imaging in Band 6, which we do not use because we prioritize the highest angular resolution imaging. \zw\ was observed twice on 2019 June 20th (maximum recoverable scale [MRS] = 0.4\arcsec) and once 2021 August 21 (MRS = 0.6\arcsec) for a total of 6649 s on source. \iras\ was observed on 2021 August 30 (MRS = 0.5\arcsec) and on 2021 August 8 in the C43-8 configuration (MRS = 0.8\arcsec) and for a total of 3160 s on source. For all measurement sets we follow the standard calibration pipeline and manually examined each measurement set to make sure flagging was complete. Then, for \iras\, we concatenated both measurement sets, checking for any inconsistencies in the relative weightings of each configuration. Compared to only using the configuration 9 measurement set\textbf{,} the combined measurement sets give an rms that is a factor of $\sim$$\sqrt{2}$ lower and improves our signal by approximately the same value because of the increased effective exposure time and uv-sampling of the slightly more compact C43-8 configuration.  Next, we imaged the line-free channels to produce continuum maps for each galaxy. This included avoiding not only the CO(2-1) line in both galaxies, but also lines like CS(5-4) and HC$_\mathrm{3}$N which are prominent in the core of \zw. We then perform continuum subtraction in the uv-plane and image the spectral window containing CO, producing a spectral cube for each galaxy. After imaging with Briggs weighting (\citealt{Briggs1995}; \verb|robust| = $-$0.5), our achieved (averaged) synthesized beam resolutions (FWHMs) for the data sets used in our analysis for \iras\ and \zw \, respectively, are 41 and 30 mas for the spectral cubes and 29 and 26 mas for the continuum images. All of these data are firmly under the resolution used in the prior dynamical measurements. For \zw, we achieved a rms of 1.11 mJy beam$^{-1}$ in a channel width, $\Delta$v, of 10.4 km s$^{-1}$ and a continuum rms of 0.033 mJy beam$^{-1}$ while for \iras \ we achieved a rms of 0.43 mJy beam$^{-1}$ in a channel width, $\Delta$v, of 11.2 km s$^{-1}$ and a continuum rms of 0.029 mJy beam$^{-1}$ . Table~\ref{table1} lists relevant properties of these measurement sets and Figure~\ref{fig:mosaics} shows Hubble I-Band images (GO-10592; PI: Evans) of both of \zw\ and \iras\ overlaid with ALMA Band 6 CO(2-1) contours. These contours were made using CO(2-1) imaged at a slightly lower resolution (\verb|robust| = 0.5) than the dataset we use for our analysis.

We use the Common Astronomy Software Applications (\textit{CASA}, \citealt{CASA2022}) package developed by the NRAO, ESO, and NAOJ to do this imaging. \textbf{tclean}, a routine within \textit{CASA}, takes the measured visibilities and reconstructs a sky model. One of the \textbf{tclean} parameters that is part of the Briggs weighting scheme (\citealt{Briggs1995}), \verb|robust|, adjusts how different parts of the uv-plane (in Fourier space) are weighted. To test the impact of the \verb|robust| parameter on our results, we performed our analysis on a range of values ($-$0.5, 0, 0.5, 2). For our galaxies using a \verb|robust| parameter of $-$0.5 recovers 90\% of the flux compared to the default pipeline value of (0.5) and provides the benefit of producing an image with a narrower beam FWHM. 

Achieving the highest available spatial resolution matches our science goal by probing gas mass as close to the black hole as possible and so we chose \verb|robust| = $-$0.5 as our default imaging for the analysis in this paper (both for CO and continuum).

\section{Morphology}
\label{sec:measurements}

\begin{table*}[t]
\centering
    
\scalebox{0.9}{

\hskip-3.4cm\begin{tabular}{c c c c c c c c}
    \hline    
         Galaxy & D$_L$ & Merger & ALMA & Resolution &  Freq. & CO Flux & Cont. Flux \\
         \ & (Mpc) & Stage & Band& (pc) & (GHz) & (Jy km s$^{-1}$) & (mJy) \\
         \hline
         \hline
          \zw\ & 110  & Pre-Merger & 3 & 59.53 &  106.88 & 4.61 & 3.67 \\
         \ & \  & \ & 6 & 13.62 &  224.38& 1.61 & 5.04 \\ 
         \hline
         \iras & 209.9  & Late-Stage & 3 & 97.15 &  104.79 & 2.33 & 2.98 \\
         \ & \ & \ & 6 & 19.23 &  219.84 & 8.79 & 2.80 \\
         \hline
    \end{tabular}}
    \caption{List of galaxy and observational properties used in this work. All data were taken in the long baseline ALMA 12m configuration. The frequency column shows the observed CO(1-0) and CO(2-1) frequencies. The merger stage definitions are from \cite{Stierwalt2013} where pre-merger is defined as galaxies prior to their first encounter and late-stage describes two nuclei in a common envelope (but that still show signs of a merger). Fluxes listed are integrated flux within the 77 mas aperture used throughout this work. Absolute flux uncertainties in ALMA Band 3 and Band 6 data are 5 and 10$\%$, respectively (\citealt{ALMAHandbook}).}
    \label{table1}
\end{table*}

\begin{figure*}[t]  
    \centering
    \includegraphics[width=16.18cm]{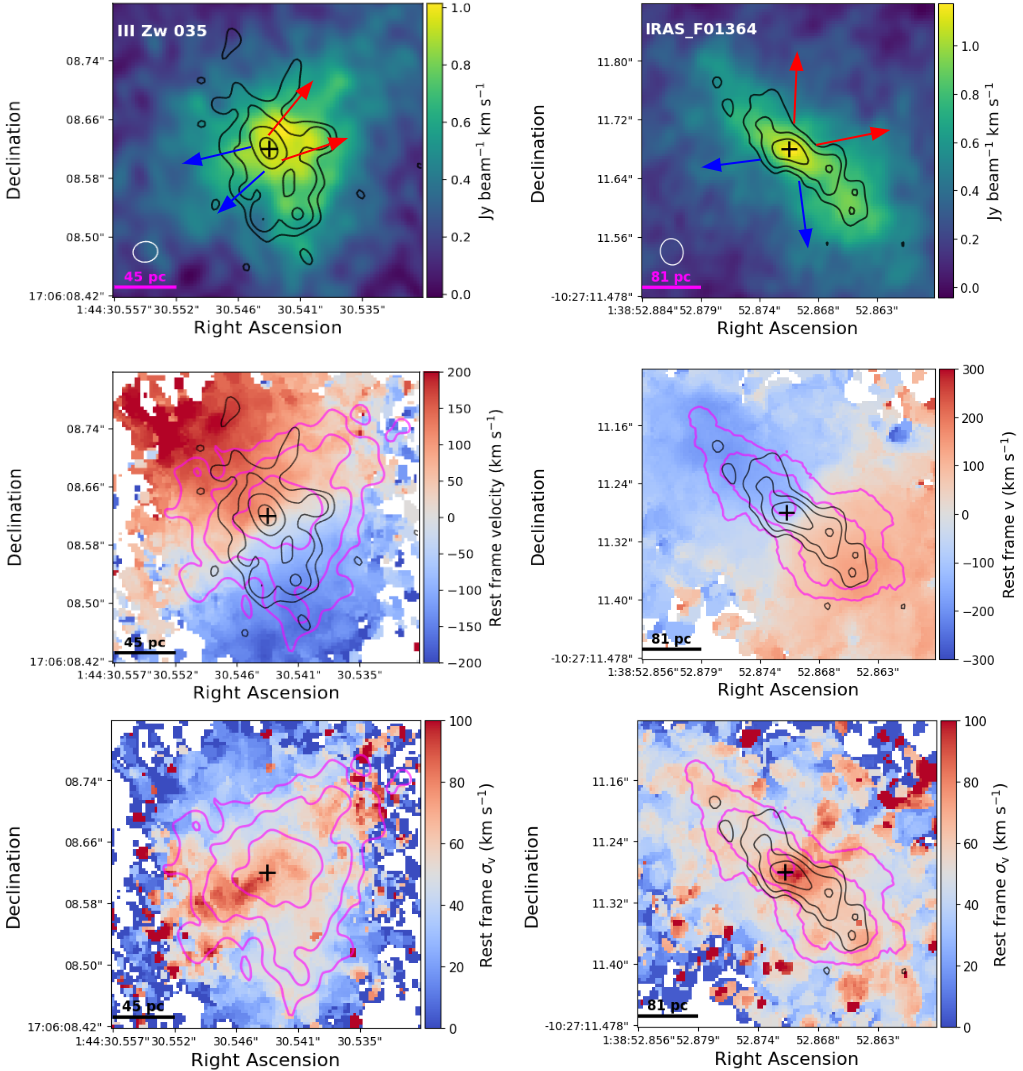}
       \caption{\textit{Top:} CO(2-1) moment 0 (integrated intensity) maps for \zw \ (left) and \iras \ (right) with Band 6 continuum contours at 230 GHz for \zw\ and 228 GHz for \iras\ with levels at 3, 6, 12, 24, 48 $\times$ (rms)  Jy beam$^{-1}$ (rms values are 32.7 $\times$ and 29.4 $\times$ $\mu$Jy beam$^{-1}$ for \zw\ and \iras\, respectively). White ellipses indicate the beam sizes of the CO images (33 $\times$ 28 for \zw\ and 43 $\times$ 40 mas for \iras. \textit{Bottom; Middle:} CO(2-1) moments 1 (velocity) and 2 (velocity dispersion) maps. Black contours are the same as in the top two panels for continuum, while magenta contours are moment 0 contours from 10$^{-0.4}$ to 10$^{0.5}$ Jy beam$^{-1}$ km s$^{-1}$ for \zw \ and 10$^{-0.8}$ to 10$^{0.5}$ Jy beam$^{-1}$ km s$^{-1}$ for \iras\ in 7 logarithmically-spaced intervals. Velocity maps for both galaxies are clipped to a 3$\sigma$ CO(2-1) detection. Continuum center, where we expect the black hole to be located, is indicated by a black cross. Red and blue arrows indicate the directionality (red and blue-shifts relative to the galaxy's systemic velocity) of the molecular outflows.}
\label{fig:COims}
\end{figure*}

\subsection{\zw}
As defined by the merger classification from \cite{Stierwalt2013} in optical wavelengths, \zw \ is a pre-merger. This means that this galaxy and its companion have not yet had their first encounter. Figure~\ref{fig:COims} shows ALMA Band 6 CO, and continuum, along with a moment 1 (velocity) map of CO(2-1) for \zw\ and \iras. The high resolution $\sim$26 mas molecular gas emission in \zw \ reveals a resolved core in CO with a peak integrated intensity that is slightly offset (5-10 pc) from the center of Band 6's continuum emission ($\sim$230 GHz). Continuum contours highlight several resolved clumpy substructures about 20-30 pc away from the central CO source. Previously, \cite{Pihlstrom2001} found clumps of OH maser and 18 cm continuum emission in similar positions around the central source. Further inspection of the ALMA data cubes in these clumps shows the presence of CS(5-4), a good dense gas tracer (\citealt{Wang2011}). The clumpy nature of this dense gas tracer and evidence from previous observations could indicate that there are star-forming clumps around \zw's core. These clumps may also be part of a forming or otherwise evolving torus or ring around \zw's nucleus, as distances of the clumps to the core are typical (10s of pc) for these sorts of structures (e.g. \citealt{GATOS2021}). CO(2-1) is not apparent in all of these clumps. Additional matched resolution data at other wavelengths, such as deeper ALMA Band 7 and Band 3 continuum, are necessary to further constrain the nature of this distribution of mass.
\par
The CO moment 1 maps (Figure~\ref{fig:COims}, bottom left) show strong disk-like rotation across the major axis of the galaxy, similar to what is seen in H$_2$ and Br$\gamma$ (\citealt{Medling2014}). \zw \ also shows evidence of a molecular outflow along the minor axis seen in the H$_2$ and Br$\gamma$ maps in \cite{2019ApJ...871..166U}, as well as in CO(1-0) at lower resolution in \cite{Lutz2020}. \zw's outflow can also be seen in CO(2-1) and it will be discussed in detail in a future work. 

\subsection{IRAS F01364-1042}
Toward the opposite end of the merger spectrum, \iras \ is classified as a late-stage merger in \cite{Stierwalt2013}. The two galaxies have their nuclei in the same envelope at the center of the system. As seen in Figure~\ref{fig:COims}, compared to \zw, the brightest CO(2-1) emission in \iras \ has an edge-on disk-like morphology that traces the continuum emission well, with a central source that seems to be marginally resolved. Similar to \zw, it shows disk-like rotation in the high-resolution CO(2-1) imaging, with symmetrical velocity gradients on either side of the disk that match well to the H$_2$ and Pa$\alpha$ maps at $\sim$2-3$\times$ coarser resolution in \cite{Medling2014}. The CO(2-1) data presented in this work show evidence of the presence of an outflow, along the minor axis, originating from the nucleus. This outflow is also supported by MIRI H$_2$ and ALMA CO(1-0) maps (Song et al. in review).

\section{Computing Nuclear Gas Mass}
\label{sec:Analysis}
Our goal in this work is to calculate the amount of the nuclear molecular gas mass that falls within the resolution limit of Keck-OSIRIS+AO ($\sim$77 mas). Following \cite{Medling2019}, in this section we describe our two main methods of calculating molecular gas mass. Additionally, we perform independent enclosed mass estimates for comparison to \cite{Medling2015} through position-velocity (PV) diagrams in
Section~\ref{sec:PVdiagrams}. We summarize and comment on the results of these analyses in Section~\ref{sec:Discussion}.
\par

As in \cite{Medling2019}, we measure nuclear gas masses to compare with \menc\ derived from stellar and gas kinematics from \cite{Medling2015}. We include these original BH masses for \zw\ and \iras\ in the top row of Table~\ref{table:gasmasses}. In this study we do not include comparisons to \zw's enclosed warm gas mass because of high uncertainties attributed to the  H$_\mathrm{2}$ and Br$\gamma$ outflow seen alone its minor axis \cite{2019ApJ...871..166U}. For \iras, we use the \menc\ derived via gas kinematics, which could be impacted by the potential outflowing component; however, no stellar dynamics-based mass is available.

\subsection{Gas mass via CO(2-1)}
\label{sec:CO21gasmass}

While H$_2$ is an abundant interstellar gas, its net-zero dipole moment makes carbon monoxide a commonly used tracer molecule for cold H$_2$ (\citealt{Solomon1972}; \citealt{Wilson1974};
\citealt{ScovilleSolomon1975}; \citealt{Burton1975}). To calculate a molecular gas mass from our spectral cube we start with the full, 1.875 GHz wide cube, then, in python, integrate across the channels with CO emission in each pixel. To determine if a channel has CO emission in it, we use a combination of visual inspection of the individual channel maps and integrated spectra in a rectangular region made in \textit{CASA}. This produces a moment 0, or flux map of CO(2-1). Then, to include broader spatial integration regions we step outward in a box shape for CO-detected pixels. Each radial step expands the box in two pixels in each x($\pm$1) and y($\pm$1) direction. The same radial stepping method is used in Section~\ref{sec:contgasmass}.
To convert to mass we must first translate the CO line flux to CO line luminosity L$_{\mathrm{CO(2-1)}}'$. For this we apply the equation from \cite{Solomon2005}:
\begin{multline}
\label{eqn:lineluminosity}
\mathrm{L}_{\mathrm{CO(2-1)}}' = (3.25 \times 10^7) \mathrm{S_{CO}\Delta v}(\nu_{\mathrm{obs}}^{-2}\mathrm{D}_L^2(1+z)^{-3}) \\
\mathrm{K \ km s}^{-1} \mathrm{pc}^2
\end{multline}
where \sco is the flux of the CO(2-1) emission (in Jy km~s$^{-1}$), $\nu_{\mathrm{obs}}$ is the observed frequency (in GHz), and D$_L$ is the luminosity distance (in Mpc).

To convert from this line luminosity to mass we use the conversion factor $\alpha _{\mathrm{CO}}$. The use of low-J CO lines as tracers of molecular gas has been explored for decades; $\alpha_{\mathrm{CO}}$ or similar conversion factors have been calibrated for both the Milky Way and other galaxies (e.g. as reviewed by \citealt{Dickman1978}, \citealt{Combes1991}, and \citealt{Bolatto2013}). This conversion factor varies depending on where it is being applied, and the inclination angle of the galaxy. For example, on $>$600 pc scales, \cite{Sandstrom2013} find mean $\alpha _{\mathrm{CO}_{(1-0)}}$ values ranging from 2.9 to 8.2. In NGC 3351, using $\sim$100 pc resolution ALMA data, \cite{Teng2022} find intensity-weighted values between 1.11 and 1.79 on 2 kpc scale, and find that the conversion factor is lower in inflow regions. \cite{Medling2019} used an $\alpha_{\mathrm{CO(2-1)}}$ conversion factor calibrated spatially by \cite{Cicone2018} for the nuclear regions of NGC 6240. NGC 6240 is a local, merging LIRG like our systems, but it does host X-ray detected AGN and a clear double nucleus, unlike our systems. 

In this work, as in \cite{Medling2019}, we use the $\alpha_{\mathrm{CO(1-0)}}$ = 2.3 $\pm$ 1.2 which is derived from the central box `C1' of NGC 6240 and provided by C. Cicone via private communication following the methods in \cite{Cicone2018}. We scale this $\alpha_{\mathrm{CO(1-0)}}$ to $\alpha_{\mathrm{CO(2-1)}}$ by independently measuring r$_{21}$ (r$_{21}$ = $\mathrm{L}'_{\mathrm{CO(2-1)}}$/$\mathrm{L}'_\mathrm{CO(1-0)}$) at 100 mas resolution in both \zw\ and \iras\ (57 and 98 pc, respectively). To measure this new r$_{21}$ we created beam-matched images by first smoothing the CO(2-1) data to the CO(1-0) data (project code 2017.1.01235.S; PI: Barcos-Mu\~{n}oz) by using \textit{CASA}'s \textbf{imsmooth} task, which performs a Fourier-based convolution to the CO(1-0) beam in which we used a Gaussian kernel. We then used the smoothed image to measure the line luminosities in the same 154 $\times$ 154 mas box used in our analysis. The new $\alpha_{\mathrm{CO(2-1)}}$ value (which includes both H$_2$ and helium) for \zw\ is calculated as:
\begin{multline}
\label{eqn:alphaco}
\alpha_{\mathrm{CO(2-1)}} = \frac{\alpha_{\mathrm{CO(1-0)}}}{\mathrm{r}_{21}} 
=  \frac{2.3 \ \pm \ 1.2}{0.88 \ \pm 0.099} \\ 
= 2.60 \ \pm \ 1.2 \ \mathrm{M}_{\odot}
(\mathrm{K \ km \ s}^{-1} \ \mathrm{pc}^2)^{-1}.
\end{multline}
and the same calculations are done for \iras\ which has a r$_{21}$ = 0.77 $\pm~0.086$ and $\alpha_{\mathrm{CO(2-1)}}$ = 3.0 $\pm~1.2 \ \mathrm{M}_{\odot}
(\mathrm{K \ km \ s}^{-1} \ \mathrm{pc}^2)^{-1}$. Uncertainties on r$_{21}$ are the result of error propagation of the ALMA Band 3 and 6 flux uncertainties of 5 and 10$\%$\textbf{,} respectively (\citealt{ALMAGuide7}). Using these $\alpha_{\mathrm{CO(2-1)}}$ estimates rather than adopting the value from \cite{Cicone2018} increases the final gas (H$_2$ and helium) mass calculated in \zw\ and \iras\ by factors of 1.1 and 1.3, respectively. 

 \cite{Montoya2023} measure a median $\alpha_{\mathrm{CO(1-0)}}$ = 1.7 $\pm$ 0.5 \msun \ in 37 ULIRGs (ultra-luminous infrared galaxies, L$_\mathrm{IR}$ $>$ 10$^{12}$\lsun) and 3 LIRGs (see also \citealt{Herrero2019}). \cite{Sandstrom2013} found that in the central kpc of 26 starburst galaxies, $\alpha_{\mathrm{CO(1-0)}}$ is a factor of $\sim$2 smaller than in the larger scale disk. Each of these studies found median $\alpha_{\mathrm{CO(1-0)}}$ values lower than the typical global value for the Milky Way (4.4~$\mathrm{M}_{\odot} (\mathrm{K \ km \ s}^{-1} \mathrm{pc}^2)^{-1}$; \citealt{Bolatto2013}). These measurements were done on larger scales than the ones we make our \menc\ measurements in, but suggest that globally-calibrated $\alpha_{\mathrm{CO(1-0)}}$ values cannot be trusted for analysis on nuclear scales. The $\alpha_{\mathrm{CO(2-1)}}$ we adopt here is therefore uncertain particularly compared to larger scale estimates of $\alpha_{\mathrm{CO(1-0)}}$, because of the lack of studies which robustly calibrate these nuclear conversion factors on $<$ 50 pc scales in LIRGs. In this work we attempt to mitigate for these uncertainties by using the highest resolution calibrated $\alpha_{\mathrm{CO(1-0)}}$ from \cite{Cicone2018} for NGC 6240, a merging (U)LIRG.  

The CO(2-1) flux profiles, which are multiplied by $\alpha_{\mathrm{CO(2-1)}}$ to obtain a mass profile, are shown in the top two rows of Figure~\ref{fig:8panel}. Inside the 77 mas Keck-OSIRIS resolution limit achieved in \cite{Medling2015} we find (3.6$ \pm$ 1.9)$\times$ $10^7$  M$_\odot$ of molecular gas mass for \zw \ and (6.7 $\pm$ 2.8) $\times$ 10$^8$ M$_\odot$ of molecular gas mass for \iras \ within 43.7 and 76.9 pc, respectively. Fractionally these values account for between $\sim$1 and $\sim$48\% (including 1$\sigma$ errors) of the previously determined \menc\ (see Table~\ref{table:gasmasses} for details).

\subsection{Gas mass via dust continuum}
\label{sec:contgasmass}

Another way of estimating the central molecular gas mass is by measuring the cospatial dust mass using our mm-wavelength continuum imaging. For LIRGs, the mm/sub-mm continuum flux densities are typically dominated by thermal dust emission (\citealt{U2012}). Continuum measurements start with the same aperture as in Section~\ref{sec:CO21gasmass} (a 154 $\times$ 154 mas square), integrating flux density as the box grows outward. We then use the calibrated dust continuum flux density ratio to H$_2$ from \cite{Scoville2016}. This calibration improves on the \cite{Scoville2015} relation used for Arp 220, but it assumes a globally-derived, mass weighted dust temperature (T$_D$) of 25~K. In the sample of ULIRGs from \cite{U2012}, the average global T$_D$ was found to be moderate at $\sim$25-45 K. Other recent studies have shown that luminosity-weighted nuclear T$_D$ can be much higher than this (e.g. \citealt{Sakamoto2021}), and so we explore a range of temperatures from 100-500 K for our primary analysis.

As our continuum-based measurements use a relation that assumes the emission is from dust, for them to be accurate we have to ensure that any contaminants to that emission are removed. Some of the continuum flux density at $\sim$230 GHz may be due to synchrotron radiation (originating from buried AGN and/or supernova remnants) and free-free emission (\citealt{Condon2016}). We can predict the contributions of these phenomena to our continuum flux by utilizing previous observations of our galaxies at lower frequencies where synchrotron and free-free emission dominate the flux, extrapolating their spectra to our frequencies. To estimate the contribution of synchrotron and free-free emission to our measured continuum flux, we use 32.5 GHz integrated fluxes from Very Large Array observations found in \cite{2017ApJ...843..117B}. Together with integrated ALMA Band 3 observations (project 2017.1.01235.S, PI Barcos-Mu\~{n}oz, see Table~\ref{table1}), we can measure the spectral index of both synchrotron and free-free contributions together between 32.5 and $\sim$100 GHz, then extrapolate the 32.5 GHz flux to ALMA Band 6 using that spectral index. For \zw\ the spectral index $\alpha$ = -0.58 $\pm\ 0.04$ and for \iras\ $\alpha$ = -0.39 $\pm\ 0.04$. Uncertainties are calculated as a combination of image noise and errors in calibration flux uncertainties for ALMA and the VLA (\citealt{ALMAGuide7}; \citealt{Partridge2016}).

To estimate the contribution of these contaminants to the Band 6 continuum, we use the $\sim$120 mas resolution, 32.5 GHz VLA data from \cite{2017ApJ...843..117B} and the spectral indices mentioned prior. We extrapolate the 32.5 GHz integrated flux to a Band 6 flux density value using:

\begin{multline}
f_{\mathrm{extra}} = f_{\mathrm{synch}} + f_{\mathrm{free-free}}
= f_{32.5\mathrm{GHz}}(\frac{\nu_{\mathrm{cont}}}{32.5\mathrm{GHz}})^{\alpha} \\ 
\end{multline}

where $\nu_{\mathrm{cont}}$ is the center of the continuum band for each galaxy in GHz, and the flux densities (f) are in Jy. The resulting contributions to our continuum flux by both synchrotron and free-free sources are 43\% and 70\% for \zw\ and \iras\ \textbf{,} respectively.

\begin{figure*}[t]
        \includegraphics[width=8.82cm]{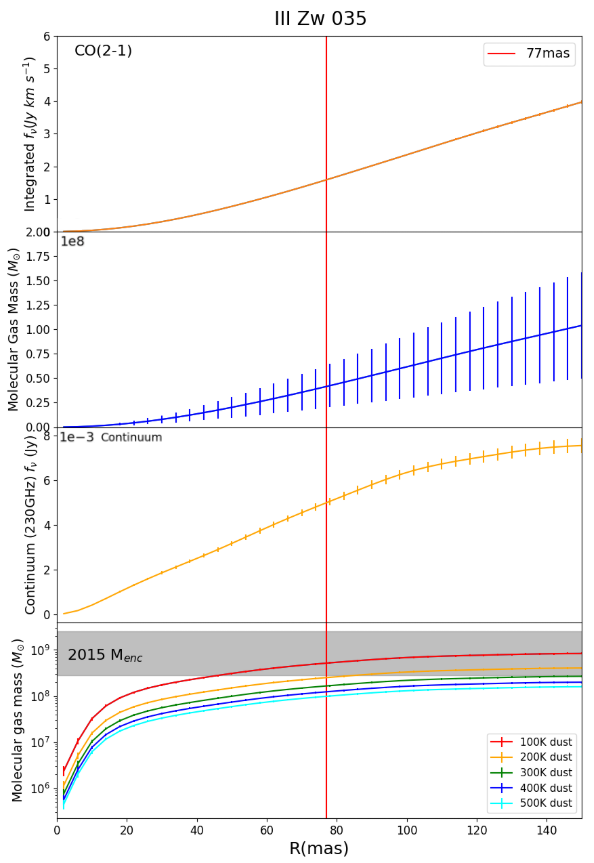}
        \includegraphics[width=8.79cm]{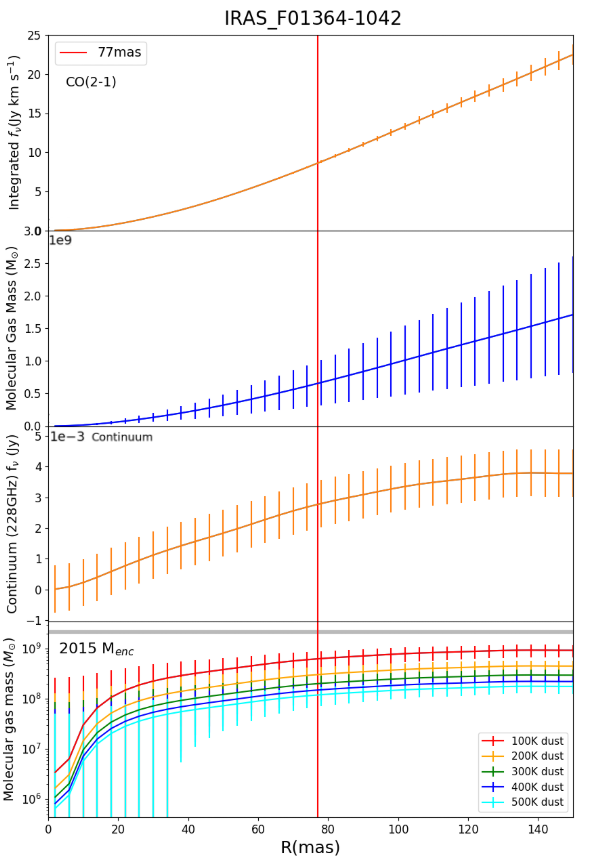}

\caption{Integrated measurements and calculated masses within the boxed aperture for \iras\ and \zw. \textit{First row}: enclosed CO(2-1) flux from images described in Section ~\ref{sec:MeasurementsandImaging}. \textit{Second row}: \menc\ calculated from CO(2-1) fluxes shown in above panels using method described in Section ~\ref{sec:CO21gasmass}. \textit{Third row}: integrated continuum flux densities at 230 GHz (\zw) and 228 GHz (\iras) extracted from images described in Section~\ref{sec:MeasurementsandImaging}. \textit{Bottom row}: \menc\ calculated from continuum fluxes shown in above panels using method described in Section~\ref{sec:contgasmass}. Grey regions are \cite{Medling2015} \menc\ ranges. \zw \ requires a T$_D$ $\gtrsim$ 175~K to have a dust-derived gas mass lower than the previous \menc\ while \iras\ requires a T$_D$ $\gtrsim$ 19~K.}
\label{fig:8panel}
\end{figure*}

\begin{table*}[t]
\centering
Integrated gas masses compared to previous \menc\ measurements
\\[0.2cm]
\begin{tabular}{l c c}
         \  & \zw\ & \iras\ \\
         \hline
         \hline
        \ \ \ \ \ \ \ \ \ \vline Gas (H$_2$, Br$\gamma$) & * & (2.2$\substack{+.060 \\ -.17}$) \ $\times$ 10$^9$ \msun  \\
         \hspace{-0.173cm}$^{(1)}$\menc\   \vline Stellar (disk) & ($>$6.8$\substack{+.10 \\ -4.0}$) \ $\times$ 10$^8$ \msun & - \\
          \ \ \ \ \ \ \ \ \ \vline Stellar (JAM) & ($<$2.0$\substack{+.50 \\ -.70}$) \  $\times$ 10$^9$ \msun & - \\
         
         \hline
         \hline
         $^{(2)}$\mgasco & (3.6 $\pm$ 1.9) $\times$ 10$^7$ \msun & (6.7  $\pm$ 2.8) $\times$ 10$^8$ \msun \\ 
         
         \hline
         $^{(4)}$Gas fraction  & $<$5.3 $\substack{+0.95 \\ -1.4}$\% & 30 $\substack{+18 \\ -18}$\%\\  
         \  & $>$1.8 $\substack{+1.0 \\ -1.1}$\%& \ \\ 
         
         \hline
         \hline
         $^{(3)}$\mgasconthundred & (5.1 $\pm$ 0.22) $\times$ 10$^8$ \msun & (6.4 $\pm$ 2.6) $\times$ $10^8$ \msun \\
         
         \hline
         $^{(4)}$Gas fraction & $>$26 $\substack{+9.0 \\ -6.5}$\% & 28 $\substack{+12 \\ -12}$\% \\ 
         \ & $<$75 $\substack{+44 \\ -3.4}$\% & \ \\
         \hline
         \hline
         $^{(3)}$\mgascontfhundred & (9.8 $\pm$ 0.43) $\times$ 10$^7$ \msun & (1.2 $\pm$ 0.50) $\times$ $10^8$ \msun \\
         
         \hline
         $^{(4)}$Gas fraction & $>$4.9 $\substack{+1.2 \\ -1.7}$\% & 5.4 $\substack{+2.2 \\ -2.3}$\% \\ 
         \ & $<$14 $\substack{+8.5 \\ -0.7}$\% & \ \\
    \end{tabular}
    \caption{Summary of results for mass estimates via CO ($\alpha_{\mathrm{CO}}$ method) and thermal dust continuum. For \zw \ we compare to \menc\ from stellar kinematics and in \iras \ we compare to \menc\ from gas kinematics. *\zw \ gas-based dynamical masses exist in \cite{Medling2015}, but we do not use them for comparison in this work due to likely contamination from a molecular outflow.
    (1): \menc\ from \cite{Medling2015}. (2): Gas mass from CO(2-1) flux as calculated in Section~\ref{sec:CO21gasmass}. (3): Gas mass from thermal excess continuum flux as calculated in Section~\ref{sec:contgasmass}. (4): Percentages displayed are relative to the kinematically-derived \menc\ found in \cite{Medling2015} and are computed as $\mathrm{M_{gas}}/\mathrm{M_{enc}}$.}
    \label{table:gasmasses} 
\end{table*}
These contributions are subtracted in the following equation, which adopts a modified blackbody spectral index $\beta$ = 1.8, where we perform our gas mass conversion (Equation 3; \citealt{Scoville2015}): 
\begin{equation}
\label{eqn:contcalc}
    \frac{0.868~\times~(\mathrm{f}_{\mathrm{cont}}-\mathrm{f}_{\mathrm{extra}})\mathrm{D}^2_{L}}{(1+z)^{4.8}\mathrm{T}_{25}\nu\substack{3.8\\350}\Gamma_{RJ}10^3\mathrm{Mpc}}10^{10}\mathrm{M_\odot}
\end{equation}

where the fluxes (f$_{\mathrm{cont}}$ is continuum flux density) are in mJy, luminosity distance, D$_\mathrm{L}$, is in Mpc, T is normalized to 25 K, $\nu$ normalized to 350 GHz, and $\Gamma _{\mathrm{RJ}}$ is the correction for departure in the rest frame of the Planck function from Rayleigh-Jeans, which varies with temperature (\citealt{Scoville2016}).

Using two bracketing dust temperature cases, 100 K and 500 K, we find gas masses inside the 77 mas Keck-OSIRIS resolution (box) limit of $(9.8 \pm\ 0.43)\ $$ \times\ 10^7$ M$_\odot$ to $(5.1 \pm 0.22) \times 10^8$ M$_\odot$ of mass for \zw \ and $(1.2 \pm 0.50) \times 10^8$ M$_\odot$ to $(6.4 \pm 2.6) \times 10^8$ M$_\odot$ of mass for \iras\ within 43.7 and 76.9 pc, respectively. Rows 3 and 4 of Figure~\ref{fig:8panel} show the continuum flux density and molecular gas masses as a function of radius. Fractionally, in relation to the previously determined \menc\ from \cite{Medling2015}, these measured values account for as little as $\sim$5\% in the high T$_D$ case to as much as $\sim$75\% in the low temperature case (see Table~\ref{table:gasmasses} for details). This wide range of gas masses and the topic of dust temperature are discussed within Section~\ref{sec:uncertaindust}.
\begin{figure*}[!ht]
    \centering
    \epsscale{0.5}  
    \includegraphics[width=0.450\linewidth]{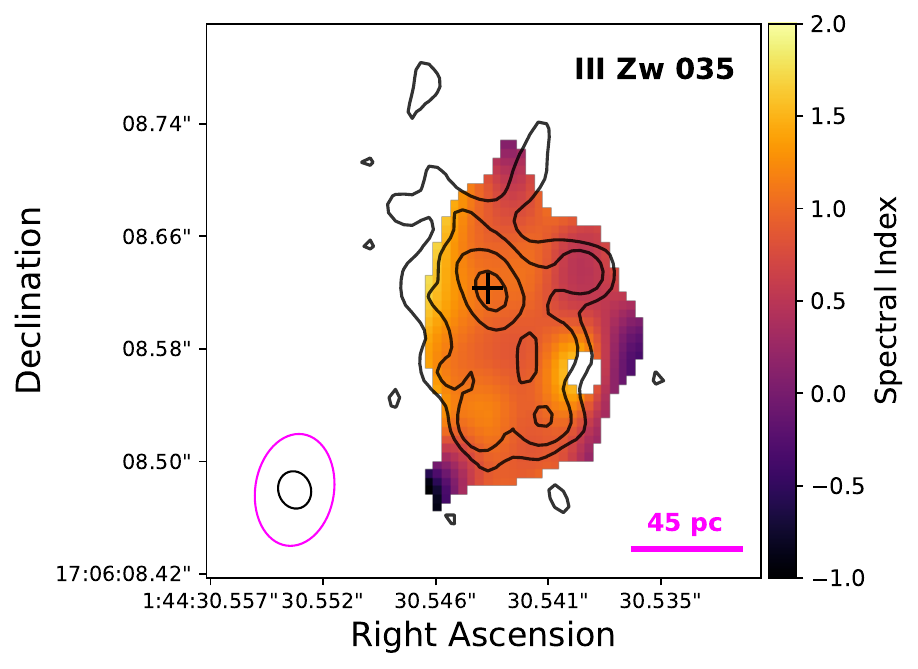}
    \includegraphics[width=0.460\linewidth]{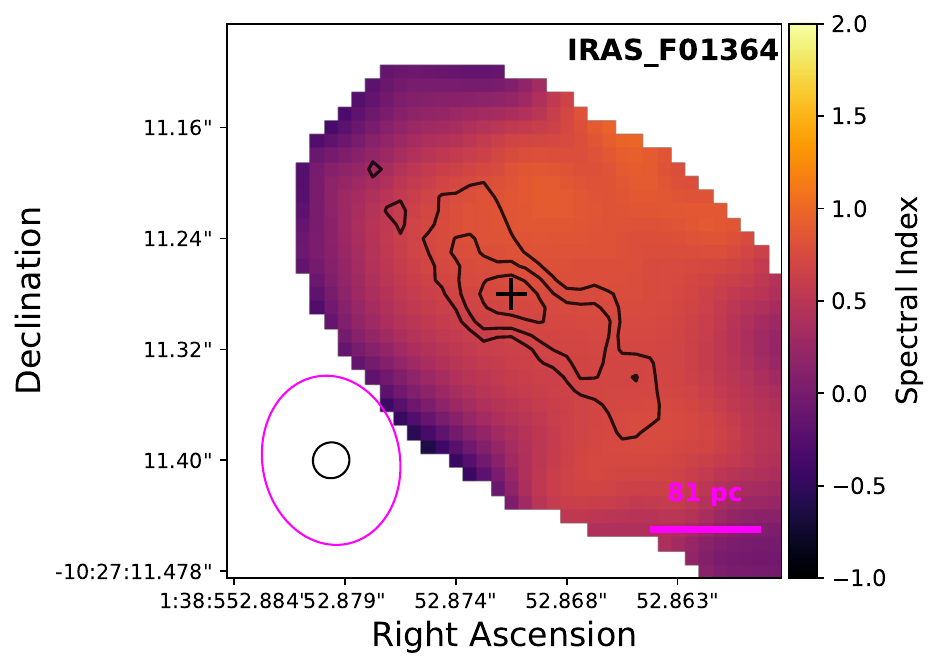}\par\medskip

    \hspace{-0.24cm}\includegraphics[width=0.462\linewidth]{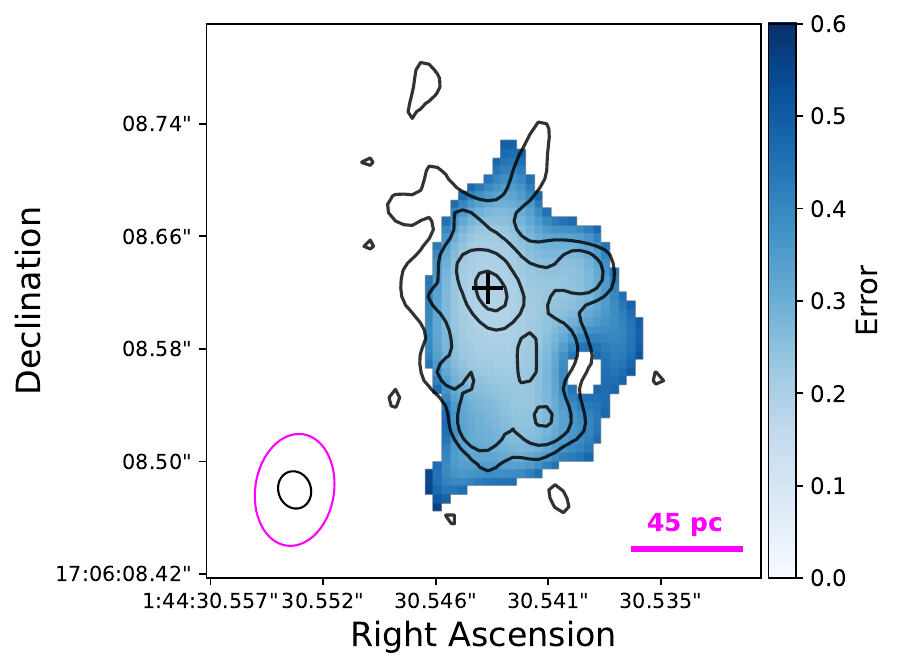}\hspace{-0.12cm}
    \includegraphics[width=0.460\linewidth]{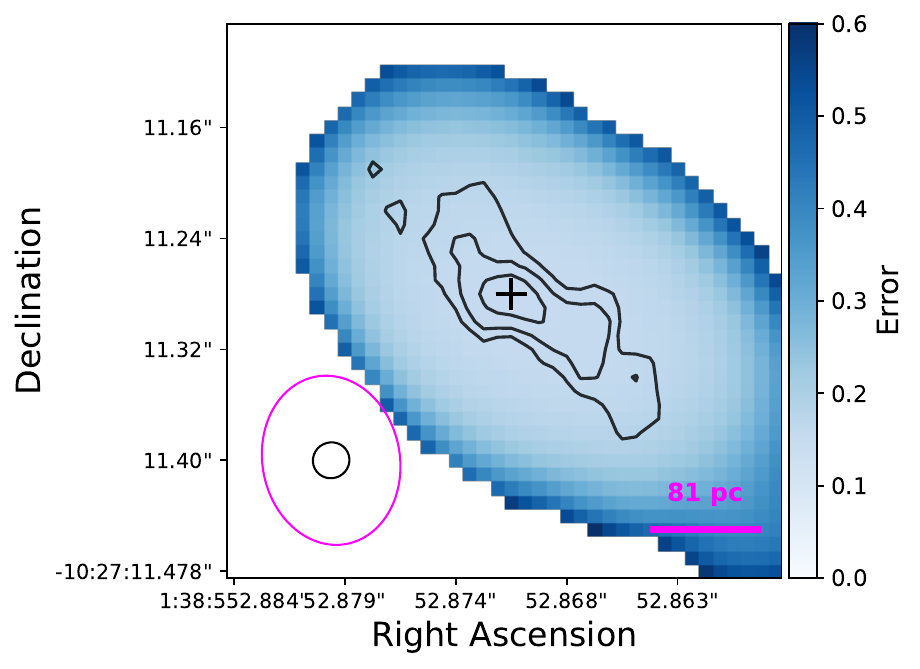}
    \caption{Spectral index (top) and uncertainty maps (bottom) for \zw\ (left) and \iras\ (right). The beam sizes of the continuum images used to create the spectral index maps (0.075\arcsec\ $\times$ 0.052\arcsec\ for \zw\ and 0.134\arcsec\ $\times$ 0.110\arcsec\ for \iras) are shown in the lower left corners in magenta. The black contours correspond to the Band 6 combined line-free continuum emission at 230 GHz for \zw\ and 228 GHz for \iras\ with levels at 3, 6, 12, 24, 48 $\times$ (rms)  Jy beam$^{-1}$ (rms values are 32.9 and 38.3 $\mu$Jy for \zw\ and \iras\, respectively). Their beam sizes are shown in the lower left corners in black (31 $\times$ 25 mas for \zw\ and 29 $\times$ 28 mas for \iras).}
    \label{fig:spectralindex}
\end{figure*}

\subsection{Spectral index mapping}
\label{sec:spectralindex}
To better understand the nature of the mm continuum emission in the nuclei of our galaxies, we create inter-band spectral index maps. To create such maps for our galaxies, we generate line-free continuum images for Band 3 (project 2017.1.01235.S, PI Barcos-Mu\~{n}oz) and Band 6 (data presented here) with the same pixel and beam sizes and using a \verb|robust| = $-$0.5. We then compute the spectral index per pixel defined as: 
\begin{equation}
\label{eqn:spindex}
\alpha = \log(\frac{f_\mathrm{B6}}{f_\mathrm{B3}})/\log(\frac{\nu_\mathrm{B6}}{\nu_\mathrm{B3}})  
\end{equation}
where f is the flux density in Jy and $\nu$ is the frequency (in GHz). B3 and B6 indicate the values of flux density and frequency of the images mentioned above. Data below a 3$\times$rms threshold is is masked and not mapped. As a general rule, a positive mm spectral index favors dust emission (\citealt{Planck2016}), while a steep negative mm spectral index indicates dominant synchrotron emission (see Section~\ref{sec:contgasmass}). We estimated the uncertainty of the spectral index via error propagation considering the uncertainty in the flux density due to noise in the image and the ALMA flux calibration uncertainties.

In Figure~\ref{fig:spectralindex}, we show the spectral index maps and related error maps for both galaxies. Given the resolution restrictions of our spectral index maps, we don’t have independent measurements of the spectral indices of the nuclei versus the clumpy torus-like structures in \zw. The integrated interband (Band 3 to 6) spectral index values for \zw\ and \iras, respectively, are 0.87 $\pm$ 0.31 and 0.49 $\pm$ 0.27. Alongside the f$_\mathrm{extra}$ measurements in Section~\ref{sec:contgasmass} that estimate contribution fractions of free-free and synchrotron emission of $<$50\%, these spectral index maps suggest that thermal dust emission is a major contributor to the Band 6 continuum emission over other non-thermal or free-free contributions.  Further observations at matched resolution to the Band 6 data are necessary to learn more about the radio SEDs of these cores and clumps to make a more direct comparison to our other results. 

\begin{figure*}[!ht]
    \centering
    \epsscale{0.5}  
    \hspace{0.95cm}
    \includegraphics[width=0.450\linewidth]{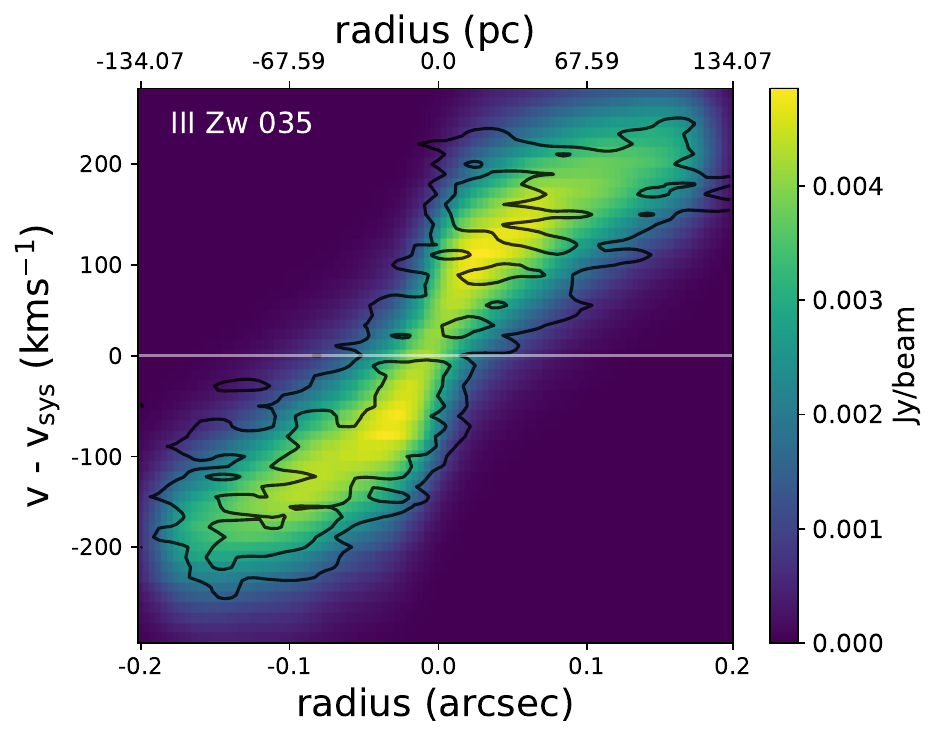}
    \hspace{-0.05cm}
    \includegraphics[width=0.450\linewidth]{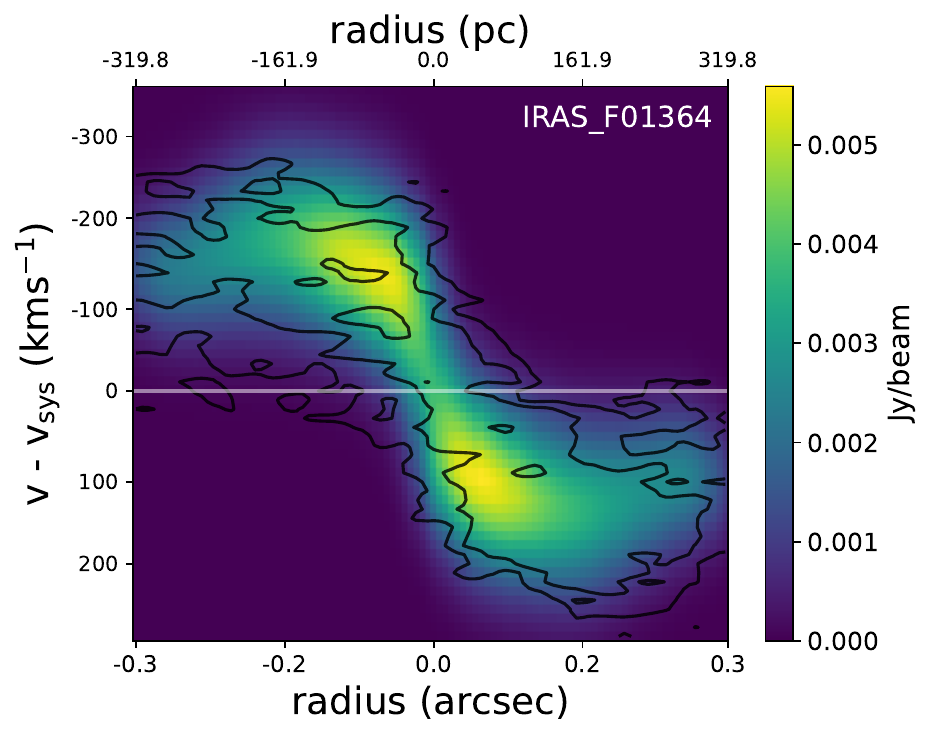}\par\medskip
    \hspace{0.25cm}
    \includegraphics[width=0.462\linewidth]{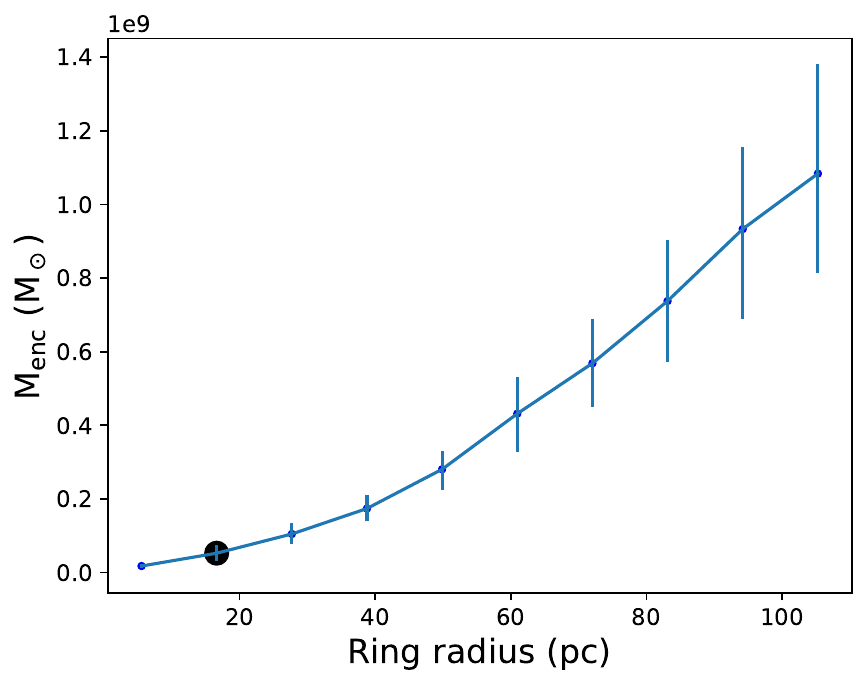}\hspace{-0.12cm}
    \includegraphics[width=0.462\linewidth]{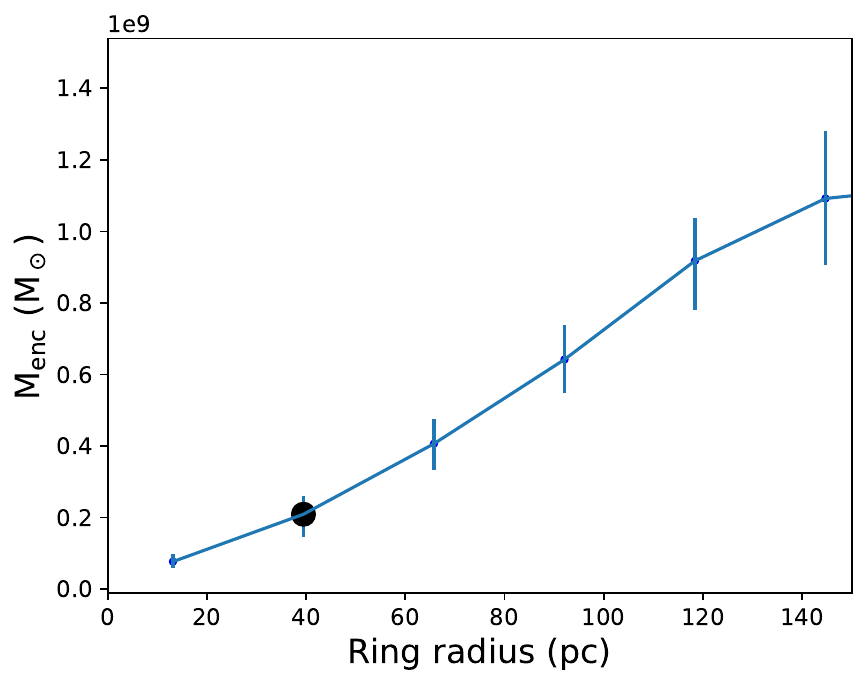}

    \caption{Position-velocity (PV) diagrams of the CO emission modeled across the major axis in \zw \ (left) \& \iras \ (right). \textit{Top:} PV models from $^{3\mathrm{D}}$Barolo outputs. Black contours correspond to 3, 6, 9, and 12 $\times$ rms of high resolution PV diagrams made in CARTA at the same central position with the same high resolution input cubes described in Section~\ref{sec:MeasurementsandImaging}. The white, horizontal lines represent velocities of zero with respect to systematic. \textit{Bottom:} Inclination-corrected \menc\ 
     calculated using the methods described in Section~\ref{sec:PVdiagrams}. \menc\ was calculated on tilted rings that were separated by single beam widths. Both \menc\  values (measured at radii of a single beam FWHM indicated by the black circles) are about an order of magnitude lower than the measurements made in \cite{Medling2015}.}
\label{fig:PVdia}
\end{figure*}

\begin{table*}[t]
\centering
Integrated gas masses compared to \menc\ measured in this work
\\[0.2cm]
\begin{tabular}{l c c}
         \ & \zw\ & \iras\ \\
         \hline
         \hline
          $^{(1)}$M$_{\mathrm{enc, CO}}$ & (5.3$\substack{+2.2 \\ -2.1}$)$\times$ 10$^7$ \msun & (2.1$\substack{+0.52 \\ -0.63}$) \ $\times$ 10$^8$ \msun  \\
         
         \hline
         \hline
         $^{(2)}$\mgasco & (8.6 $\pm$ 4.1) $\times$ 10$^6$ \msun & (2.4  $\pm$ 1.0) $\times$ 10$^8$ \msun \\ 
         
         \hline
         $^{(4)}$Gas fraction & 16 $\substack{+10 \\ -10}$\% & 110 $\substack{+56 \\ -59}$\%\\  
         
         \hline
         \hline
         $^{(3)}$\mgasconthundred & (1.9 $\pm$ 0.98) $\times$ 10$^8$ \msun & (3.2 $\pm$ 2.6) $\times$ $10^8$ \msun \\ 
                 
         \hline
         $^{(4)}$Gas fraction  & 370 $\substack{+160 \\ -150}$\% & 150 $\substack{+130 \\ -130}$\% \\ 
         
         \hline
         \hline
         $^{(3)}$\mgascontfhundred & (3.7 $\pm$ 0.19) $\times$ 10$^7$ \msun  & (6.2 $\pm$ 5.0) $\times$ $10^7$ \msun \\ 
         
         \hline
         $^{(4)}$Gas fraction & 70 $\substack{+30 \\ -28}$\% & 29 $\substack{+25 \\ -26}$\% \\ 
         
    \end{tabular}
    \caption{Summary of results for mass estimates via CO ($\alpha_{\mathrm{CO}}$ method) and thermal dust continuum when subtracted from \menc\ calculated in this work (see Section~\ref{sec:PVdiagrams}). All values in this table are measured at a radius equal to the beam size of the respective CO data. (1): \menc\ derived using CO(2-1) kinematics in Section~\ref{sec:PVdiagrams}. (2): Gas mass from CO(2-1) flux as calculated in Section~\ref{sec:CO21gasmass}. (3): Gas mass from continuum thermal excess flux as calculated in Section~\ref{sec:contgasmass}. (4): Percentages displayed are relative to \menc\ derived from CO kinematics and are calculated as $\mathrm{M_{gas}}/\mathrm{M_{enc, CO}}$.}
    \label{table:gasmassescoldgas}
\end{table*}

\section{Independent enclosed mass measurements}
\label{sec:PVdiagrams}
The results in Section~\ref{sec:Analysis} present corrected \cite{Medling2015} \mbh\ values for \zw\ and \iras. With those corrections, the black holes remain overmassive except in the case of \zw\ with T$_D$ $\lesssim$ 175 K. In this subsection, we independently measure \menc\ at the Band 6 beam resolution using the CO cubes presented in this work.

\subsection{Tilted-ring modeling}
\label{sec:tiltedrings}
To calculate dynamical \menc\, we model the CO kinematics using the tilted-ring modeling algorithm $^{3\mathrm{D}}$Barolo (\citealt{3DBarolo}). Using our high resolution CO(2-1) cubes, $^{3\mathrm{D}}$Barolo models tilted rings with inclination-corrected rotational velocities to the line emission at a range of distances from the center. This 3D modeling approach is preferred over a 2D method in large part due to the instrumental effect of beam smearing being accounted for during the convolution step (\citealt{3DBarolo}). We provide initial guesses to $^{3\mathrm{D}}$Barolo's two-stage \verb|3DFIT| task for parameters like inclination, PA, and redshift. We tested different methods of building the $^{3\mathrm{D}}$Barolo mask as well, using both SEARCH and SMOOTH\&SEARCH. Models and residuals produced in this way are shown in Appendix~\ref{app:kinematic_modeling}. In \zw, the outflow likely dominates the velocity dispersion map. Position angles modeled by $^{3\mathrm{D}}$Barolo are about 15 degrees lower than those in \cite{Medling2014}. This difference in PA may be due to the modeling in this work using kinematics of the cold gas rather than warm gas.

We then use the inclination-corrected, circular velocities v at radii R of the kinematic models computed by $^{3\mathrm{D}}$Barolo in our \menc\ calculation,  $\mathrm{M}_{\mathrm{enc}}(\mathrm{R}) = \mathrm{v}^2 \mathrm{R}\mathrm{G}^{-1}$. This method assumes a spherical mass distribution. The \menc\ profiles, along with the position-velocity models from $^{3\mathrm{D}}$Barolo and contours of position-velocity diagrams manually made in CARTA (Cube Analysis and Rendering Tool for Astronomy, \citealt{CARTA}), can be seen in Figure~\ref{fig:PVdia}.

\subsection{Comparison to warm gas M$_\mathrm{enc}$}

\begin{figure*}
    \centering
    \includegraphics[width=1\linewidth]{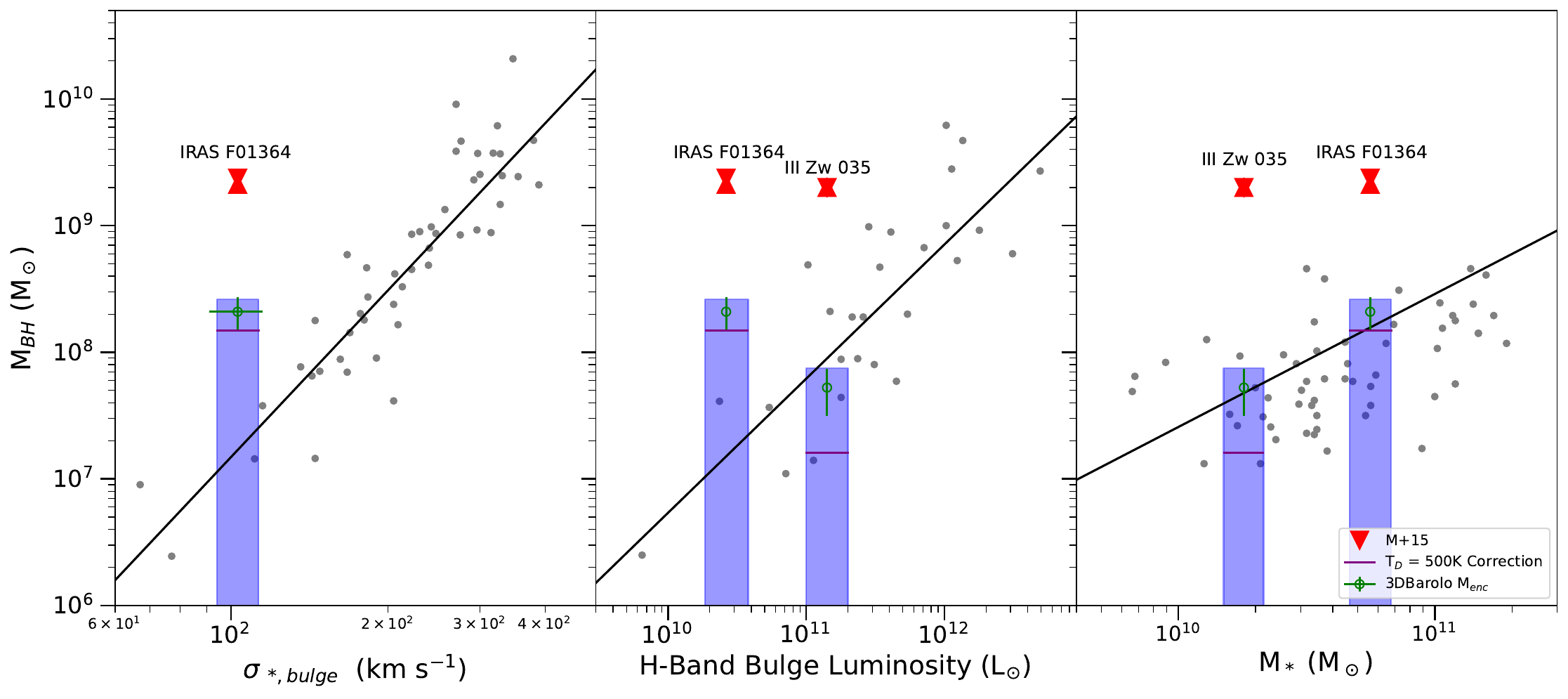}
    \caption{\cite{Medling2015} (red) and Section~\ref{sec:PVdiagrams} (green) \menc\ plotted on black hole scaling relations: \mbh\ vs. $\sigma_\star$ (left), bulge luminosity L$_{\mathrm{bulge}}$ (middle), and total stellar mass M$_\star$ (right). Grey points and line fits are based on various literature measurements, see Section~\ref{sec:gasmassincontext} for details. Blue shaded regions represent the widest range of molecular gas-corrected \mbh\ (M$_\mathrm{enc}$ - M$_\mathrm{gas}$) where M$_\mathrm{enc}$ is the value computed in this work (see Section~\ref{sec:PVdiagrams}). Both the CO-derived gas value and the T$_D$ = 100-500 K continuum methods are included in this range. The purple horizontal lines are the continuum-corrected values with T$_D$ = 500K. In all cases, the enclosed gas mass corrects the total enclosed mass measured in this work down to (or below) scaling relations.}
\label{fig:scalingrelations}
\end{figure*}

At 30 and 39 mas for \zw\ and \iras\, respectively, we find M$_{\mathrm{enc}}$ of (5.26$\substack{+2.21 \\ -2.10})\ \times$ 10$^7$ and (2.09$\substack{+0.52 \\ -0.63})\ \times$ 10$^8$ \msun (see Table~\ref{table:gasmasses}) which are  between 91-93$\%$ lower than those found in \cite{Medling2015} at 77 mas resolution. In 
Table~\ref{table:gasmassescoldgas}, we show new upper limits for the central \mbh\ by subtracting matched-resolution molecular gas masses from those high-resolution M$_{\mathrm{enc}}$. Corrected \mbh\ are also compared to BH-galaxy scaling relations in Figure~\ref{fig:scalingrelations}.

The first main difference that may be leading to different \menc\ results between the modeling here and in \cite{Medling2015} is the nature of the kinematic tracer. In this work, we use CO(2-1), which traces the cold gas in the nuclei of these galaxies while for \iras\ \cite{Medling2015} modeled the warm-gas tracer Pa$\alpha$. Given the indication of outflows present in \iras, Pa$\alpha$ may be an unreliable tracer of the bulk motion of the gas (e.g. in \citealt{Davies2024}). In \zw, although \cite{Medling2015} derived their masses from stars, which should not be impacted by outflows, the resolution of our cold gas dynamics is higher than obtained with Keck/OSIRIS. The CO beam sizes are 1.5-2 times smaller than the Keck/OSIRIS PSF. This means that we are physically probing kinematics closer to the black hole, thus presumably including less mass other than the black hole (stars, dust, gas) in the process.

The \menc\ presented in this work have the benefits of using a tracer of the dynamically cold gas and physically probing closer to the SMBH than the measurements of \cite{Medling2015}. Like the \cite{Medling2015} warm gas and stellar measurements though, for the purpose of comparison to scaling relations, these are still measurements of enclosed mass (from which we can subtract gas mass) rather than \mbh.

\section{Discussion}
\label{sec:Discussion}

\subsection{Black hole masses in context}
\label{sec:gasmassincontext}
Figure~\ref{fig:scalingrelations} shows the black hole masses with our calculated nuclear gas contaminations removed on three scaling relations: \mbh \ vs. $\sigma_\star$, bulge luminosity L$_\mathrm{bulge}$, and total stellar mass M$_\star$. For \mbh \ vs. $\sigma_\star$ we use dynamical data on elliptical and classical bulge galaxies compiled in \cite{Kormendy2013} with their equation 7 for the best-fit line. For \mbh\ vs. \ L$_\mathrm{bulge}$ we use luminosities from \cite{MarconiHunt2003} and updated \mbh\ from \cite{McConnellMa2013} with an updated fit used in \cite{Medling2019}. For \mbh \ vs. total stellar mass M$_\star$ we use data from \cite{Bennert2011} and \cite{Cisternas2011a}, along with a best-fit line from \cite{McConnellMa2013}. 

Results depend on the initial \menc\ value used for these two nuclei. If we assume that the \cite{Medling2015} \menc\ (at coarser resolution) is realistic, then the only case where we find enough molecular gas mass to shift either nucleus' dynamically-derived \mbh\ down to scaling relations is where \zw's dust temperature T$_D$$\lesssim175$ K.

On the other hand, if we assume the dynamical modeling from Section~\ref{sec:PVdiagrams} is better constrained, we find nearly the opposite. In all cases, subtracting the central molecular gas from the dynamical \menc\ causes these black holes to fall on to scaling relations. 

This dramatic distinction between the two methods could be caused by several factors. As is posited in \cite{Medling2015}, non-circular motion could have an impact on those original BH mass measurements. Warm gas is more likely to trace the turbulent ouflow than cold gas, which may have driven \textbf{up} the M$_\mathrm{enc}$ from warm gas. We find evidence for outflowing components along the minor axis in \zw\ and \iras\ in CO(2-1) at these small spatial scales. We also find evidence for mild disk warping in both \zw\ and \iras\ in CO(2-1) moment maps, although primarily beyond the radius used to calculate M$_{\mathrm{enc}}$. The disk warping properties on the smallest scales remain unknown, and incorporating corrections for these kinematic deviations -- especially for strong central inclination angle shifts -- could revise both of the \menc\ measurements. Future work focused on these galaxies will provide additional information on the magnitude of the non-circular motions caused by the outflows, which we can use to constrain their impact on these \menc\  measurements (Song et al. in review for \iras).

\subsection{Uncertain Dust Temperature}
\label{sec:uncertaindust}
For our two galaxies, as with the gas, we do not have an accurate estimate for the dust temperature on the scales studied in this work. Our analysis uses the empirical calibration for T$_D$ in ULIRGs from \cite{Scoville2016} in which gas mass scales inversely with temperature. \cite{Scoville2016} advocates for a mass-weighted T$_\mathrm{D}$ of 25 K based largely on Herschel observations of nearby galaxies (\citealt{Dunne2011}; \citealt{Dale2012}; \citealt{Auld2013}). Other studies in ULIRGs (e.g. \citealt{Sakamoto2021} and \citealt{2022ApJ...927...21W}) have shown that submm luminosity-weighted T$_D$ can exceed 500 K in their nuclei. However, luminosity-weighted T$_D$ are always higher than their mass-weighted counterparts for which the \cite{Scoville2016} relation is calibrated.

This unknown presents a challenge for the results of our continuum-based measurements. As is shown in the bottom panel of Figure~\ref{fig:8panel}, depending on the temperature assumption, the continuum-estimated molecular gas mass has a $\sim$1.5 dex range for each galaxy. At low T$_D$, at both comparison radii, the corresponding M$_\mathrm{gas}$ is driven \textit{above even} the kinematic measurements from \cite{Medling2015}. For \zw \ this temperature is $\sim$175 K and for \iras \ it is 19 K. If gas is the entirety of the \menc\ in either case, it leaves no room for a SMBH or stellar component, so we consider these T$_D$ values  to be firm lower limits. As such, our continuum-based measurements are acting as an upper limit to the molecular gas mass content, and we must wait for more well-constrained T$_D$ estimates for these galaxies to better understand the relationship between our measurement methods. CO excitation diagrams or radiative transfer modelling such as RADEX \citep{vandertak2007} at matched spatial scales with the assumption that the molecular gas and dust is well mixed (see \citealt{Viti2014} for an example in NGC 1068) could lead to such results.

\subsection{Implications for physics in nuclei of gas-rich mergers}

Whether or not these black holes still remain overmassive with respect to scaling relations depends on which \menc\ is correct. While there is evidence for an outflowing component in these systems, the physically cold gas with low turbulence is connected to dynamically cold kinematics. Because of the higher resolution of these ALMA CO data, we expect that the CO-derived M$_\mathrm{enc}$ are more reliable than previous estimates. In this case, the new enclosed masses calculated in Section~\ref{sec:PVdiagrams} are upper limits on \mbh, and both M$_\mathrm{enc}$ and M$_\mathrm{enc}$ - M$_\mathrm{gas}$ fall along all scaling relations.

If we adopt values from \cite{Medling2015}, those M$_\mathrm{enc}$ or M$_\mathrm{enc}$ - M$_\mathrm{gas}$ values all lie significantly above the scaling relations except for III Zw 035 when using a continuum-estimated M$_\mathrm{gas}$ with T$_D$ $\lesssim$175 K. As was also found in \cite{Medling2019}, kinematically-derived \mbh\ upper limits in ULIRGs may be significantly elevated due to the unresolved (gaseous) mass surrounding SMBH. The impact of cold gas contamination will depend on the nature of each individual system. In the case of the two LIRGs studied here, the fraction of the  \mbh\ estimates from \cite{Medling2015} that can be attributed to cold gas could be rather insignificant (\textbf{1}\%) to very influential (75\% or more depending on T$_D$ and comparison value). 
\par
Evolutionarily for \zw, \iras, and NGC 6240N, overmassive SMBHs would suggest a model where black hole accretion occurs before growth on the larger galaxy scale. Simulations of this growth process suggest the opposite (\citealt{Hopkins2012}; \citealt{Cen2012}; \citealt{Angles2017}). In merger-driven accretion, material has to travel to the nucleus, losing angular momentum along the way. On this path, there is expected to be a period of mixing and subsequent starburst that could cause global properties like $\sigma_\star$, L$_{\mathrm{bulge}}$, and M$_\star$ to increase before the inflowing gas reaches the nucleus and accretes onto the SMBH. Other than gas accretion, BH-BH mergers can play a part in SMBH mass build-up. These events, however, have uncertain timescales and are not expected to contribute the majority of mass gain for SMBHs like ours (e.g. in \citealt{Treister2012}; \citealt{Porras2025}). To explain a model where \mbh\ can outpace stellar growth in a merger, an angular momentum dissipation mechanism to facilitate rapid accretion of gas needs to be present.

The measurements presented here constrain \mbh\ corrections due to nuclear gas content to a lower limit. The current sample of four total nuclei (NGC6240 N/S, \zw, and \iras) is limited to these gas-rich mergers and is therefore still not representative or statistically large enough to make corrections to any general sample of SMBHs. In gas-rich LIRG mergers, at least a few to a few tens of percent of the \textbf{r $\lesssim$ 77 pc} dynamically-measured M$_\mathrm{enc}$ could be molecular gas contaminating the \mbh\ measurements.  Galaxy models should incorporate black hole mass and the full distribution of nuclear gas to properly simulate accretion physics. 

Large nuclear gas reservoirs (as are shown to form in gas-rich mergers) are likely to form a viscous accretion disk. Viscous accretion disks transport mass much slower than predicted by Bondi-like accretion models, which are theoretically predicated on free-fall of gas onto the SMBH (\citealt{Hoyle1939}, \citealt{1944MNRAS.104..273B}, \citealt{Bondi1952}, \citealt{1979SvA....23..201B}, \citealt{Mayer2007}, \citealt{Power2011}). We predict that models of gas-rich mergers that use a Bondi-like accretion prescription are overestimating accretion rates. Galactic nuclei can coalesce quicker than the larger scale disks, on timescales as short as 10 Myr (\citealt{Khan2016}).

\section{Conclusions}
\label{sec:Conclusions} 
In this work we used high resolution (sub-30 mas) CO(2-1) and continuum observations of two nearby LIRGs, \zw\ and \iras, to measure molecular gas mass within the central few 10s of pc. Fractionally, we find that between 1 and 75\% of the \menc\ calculated in \cite{Medling2015} from stellar or gas kinematics can be attributed to molecular gas within 10s of parsecs from the central black holes. At higher resolution, the \menc\  within the inner 40 mas are 91-93$\%$ lower than from kinematics taken over a larger region. Molecular gas mass contributes at least 15\% of the new \menc.

Because of the higher resolution and less kinematically disturbed nature of cold gas, we expect the new dynamically-derived \menc\ calculated in Section~\ref{sec:PVdiagrams} to be closer to the true \mbh\ of these merging systems. In all cases, starting from this \menc\, \zw\ and \iras\ fall on the \mbh\ scaling relations shown in Figure~\ref{fig:scalingrelations}. In most cases this is true even before subtracting the enclosed molecular gas mass. These new measurements are still limited in that they are enclosed masses rather than black hole masses, and our sample size is low. Very high resolution observations of cold gas for the remaining sample of overmassive black holes found in \cite{Medling2015} (IRAS F17207-0014, NGC 2623, and CGCG 436-030) and independent \menc\ measurements of NGC 6240's nuclei would allow us to understand if these nuclei \textbf{are all consistent with} scaling relations when using cold gas as a tracer.

The accuracy of accretion modeling is directly limited by our understanding of typical nuclear gas masses and distributions. With significant nuclear molecular gas, a viscous accretion disk is likely to form, causing a slower rate of accretion.  However, the resulting non-spherical accretion could potentially circumvent the Eddington limit; if super-Eddington accretion were common, it would have significant implications for the growth of supermassive black holes in the early universe and throughout cosmic time. Many cosmological simulations rely on spherically-symmetric black hole accretion rate prescriptions to predict the impacts of black hole growth and subsequent AGN feedback. The presence of nuclear gas disks could limit the power of such predictions. 

Our current sample of four nuclei (NGC6240 N/S, \zw, and \iras) is not large enough nor representative of all gas-rich mergers -- we need more high resolution datasets to push towards a unified picture of nuclear gas mass within galactic nuclei. Regardless, gas should not be left unconsidered when using dynamics to derive \mbh\ in gas-rich galaxies. 

\section{Acknowledgements}
The authors thank first the indigenous people of Hawai`i for the opportunity to be guests on your sacred mountain. We recognize the cultural significance that Maunakea holds for your community and are privileged to use data borne from science conducted in such a setting. We wish to pay respect to the Atacameño community of the Chajnantor
Plateau, whose traditional home now also includes the ALMA observatory.

We would like to thank the anonymous reviewer for their contributions to this project, in particular with regards to our independent upper limit on \mbh. This work makes use of the following data from ALMA: projects 2018.1.01123.S and 2019.1.00811.S (PI: Medling); project 2017.1.01235.S (PI: Barcos-Mu\~{n}oz). ALMA is a partnership of ESO (representing its member states), NSF (USA) and NINS (Japan), together with NRC (Canada) and NSC and ASIAA (Taiwan) and KASI (Republic of Korea), in cooperation with the Republic of Chile. The Joint ALMA Observatory is operated by ESO, AUI/NRAO and NAOJ. The National Radio Astronomy Observatory is a facility of the National Science Foundation operated under cooperative agreement by Associated Universities, Inc. Some of the data presented herein were obtained at the W. M. Keck Observatory, which is operated as a scientific partnership among the California Institute of Technology, the University of California and the National Aeronautics and Space Administration. The Observatory was made possible by the generous financial support of the W. M. Keck Foundation. The authors also wish to thank the W.M. Keck Observatory staff for their efforts on the OSIRIS+AO instrumentation. JA thanks the staff at NRAO Charlottesville for their generous mentorship during invaluable in-person visits and their assistance virtually. JA also thanks Enrico di Teodoro for his input on final model fits. JA acknowledges support from NRAO Student Observing Support program awards SOSPA7-017 and SOSPADA-017. JA and AMM also acknowledge support from NSF CAREER number 2239807. JA, AMM, and VU acknowledge partial funding support from the NASA Astrophysics Data Analysis Program (ADAP) grant number 80NSSC23K0750. VU further acknowledges partial funding support from NASA Astrophysics Data Analysis Program (ADAP) grant number 80NSSC20K0450, Space Telescope Science Institute grants, numbers HST-AR-17063.005-A, HST-GO-17285.001, and JWST-GO-01717.001. CC acknowledges funding from the European Union's Horizon Europe research and innovation programme under grant agreement No. 101188037 (AtLAST2). Some of the data presented in this article was obtained from the Mikulski Archive for Space Telescopes (MAST) at the Space Telescope Science Institute. The specific observations analyzed can be accessed via \dataset[https://doi.org/10.17909/h5ts-qy16]{https://doi.org/10.17909/h5ts-qy16}. JM is funded by the Hirsch Foundation. CR acknowledges support from the Fondecyt Iniciacion grant 11190831 and ANID BASAL project FB210003. The Flatiron Institute is supported by the Simons Foundation. CR acknowledges support from Fondecyt Regular grant 1230345 and ANID BASAL project FB210003.

\textit{Software:} Astropy \citep{astropy:2013, astropy:2018, astropy:2022}, Matplotlib (\citealt{Matplotlib}), NumPy (\citealt{harris2020array}), CASA (\citealt{CASA2022}), and $^{3\mathrm{D}}$Barolo (\citealt{3DBarolo}). 

\pagebreak
\bibliographystyle{apj}
\bibliography{paper1citations}

@ARTICLE{EHT2019,
       author = {{Event Horizon Telescope Collaboration} and {Akiyama}, Kazunori and {Alberdi}, Antxon and {Alef}, Walter and {Asada}, Keiichi and {Azulay}, Rebecca and {Baczko}, Anne-Kathrin and {Ball}, David and {Balokovi{\'c}}, Mislav and {Barrett}, John and {Bintley}, Dan and {Blackburn}, Lindy and {Boland}, Wilfred and {Bouman}, Katherine L. and {Bower}, Geoffrey C. and {Bremer}, Michael and {Brinkerink}, Christiaan D. and {Brissenden}, Roger and {Britzen}, Silke and {Broderick}, Avery E. and {Broguiere}, Dominique and {Bronzwaer}, Thomas and {Byun}, Do-Young and {Carlstrom}, John E. and {Chael}, Andrew and {Chan}, Chi-kwan and {Chatterjee}, Shami and {Chatterjee}, Koushik and {Chen}, Ming-Tang and {Chen}, Yongjun and {Cho}, Ilje and {Christian}, Pierre and {Conway}, John E. and {Cordes}, James M. and {Crew}, Geoffrey B. and {Cui}, Yuzhu and {Davelaar}, Jordy and {De Laurentis}, Mariafelicia and {Deane}, Roger and {Dempsey}, Jessica and {Desvignes}, Gregory and {Dexter}, Jason and {Doeleman}, Sheperd S. and {Eatough}, Ralph P. and {Falcke}, Heino and {Fish}, Vincent L. and {Fomalont}, Ed and {Fraga-Encinas}, Raquel and {Freeman}, William T. and {Friberg}, Per and {Fromm}, Christian M. and {G{\'o}mez}, Jos{\'e} L. and {Galison}, Peter and {Gammie}, Charles F. and {Garc{\'\i}a}, Roberto and {Gentaz}, Olivier and {Georgiev}, Boris and {Goddi}, Ciriaco and {Gold}, Roman and {Gu}, Minfeng and {Gurwell}, Mark and {Hada}, Kazuhiro and {Hecht}, Michael H. and {Hesper}, Ronald and {Ho}, Luis C. and {Ho}, Paul and {Honma}, Mareki and {Huang}, Chih-Wei L. and {Huang}, Lei and {Hughes}, David H. and {Ikeda}, Shiro and {Inoue}, Makoto and {Issaoun}, Sara and {James}, David J. and {Jannuzi}, Buell T. and {Janssen}, Michael and {Jeter}, Britton and {Jiang}, Wu and {Johnson}, Michael D. and {Jorstad}, Svetlana and {Jung}, Taehyun and {Karami}, Mansour and {Karuppusamy}, Ramesh and {Kawashima}, Tomohisa and {Keating}, Garrett K. and {Kettenis}, Mark and {Kim}, Jae-Young and {Kim}, Junhan and {Kim}, Jongsoo and {Kino}, Motoki and {Koay}, Jun Yi and {Koch}, Patrick M. and {Koyama}, Shoko and {Kramer}, Michael and {Kramer}, Carsten and {Krichbaum}, Thomas P. and {Kuo}, Cheng-Yu and {Lauer}, Tod R. and {Lee}, Sang-Sung and {Li}, Yan-Rong and {Li}, Zhiyuan and {Lindqvist}, Michael and {Liu}, Kuo and {Liuzzo}, Elisabetta and {Lo}, Wen-Ping and {Lobanov}, Andrei P. and {Loinard}, Laurent and {Lonsdale}, Colin and {Lu}, Ru-Sen and {MacDonald}, Nicholas R. and {Mao}, Jirong and {Markoff}, Sera and {Marrone}, Daniel P. and {Marscher}, Alan P. and {Mart{\'\i}-Vidal}, Iv{\'a}n and {Matsushita}, Satoki and {Matthews}, Lynn D. and {Medeiros}, Lia and {Menten}, Karl M. and {Mizuno}, Yosuke and {Mizuno}, Izumi and {Moran}, James M. and {Moriyama}, Kotaro and {Moscibrodzka}, Monika and {M{\"u}ller}, Cornelia and {Nagai}, Hiroshi and {Nagar}, Neil M. and {Nakamura}, Masanori and {Narayan}, Ramesh and {Narayanan}, Gopal and {Natarajan}, Iniyan and {Neri}, Roberto and {Ni}, Chunchong and {Noutsos}, Aristeidis and {Okino}, Hiroki and {Olivares}, H{\'e}ctor and {Ortiz-Le{\'o}n}, Gisela N. and {Oyama}, Tomoaki and {{\"O}zel}, Feryal and {Palumbo}, Daniel C.~M. and {Patel}, Nimesh and {Pen}, Ue-Li and {Pesce}, Dominic W. and {Pi{\'e}tu}, Vincent and {Plambeck}, Richard and {PopStefanija}, Aleksandar and {Porth}, Oliver and {Prather}, Ben and {Preciado-L{\'o}pez}, Jorge A. and {Psaltis}, Dimitrios and {Pu}, Hung-Yi and {Ramakrishnan}, Venkatessh and {Rao}, Ramprasad and {Rawlings}, Mark G. and {Raymond}, Alexander W. and {Rezzolla}, Luciano and {Ripperda}, Bart and {Roelofs}, Freek and {Rogers}, Alan and {Ros}, Eduardo and {Rose}, Mel and {Roshanineshat}, Arash and {Rottmann}, Helge and {Roy}, Alan L. and {Ruszczyk}, Chet and {Ryan}, Benjamin R. and {Rygl}, Kazi L.~J. and {S{\'a}nchez}, Salvador and {S{\'a}nchez-Arguelles}, David and {Sasada}, Mahito and {Savolainen}, Tuomas and {Schloerb}, F. Peter and {Schuster}, Karl-Friedrich and {Shao}, Lijing and {Shen}, Zhiqiang and {Small}, Des and {Sohn}, Bong Won and {SooHoo}, Jason and {Tazaki}, Fumie and {Tiede}, Paul and {Tilanus}, Remo P.~J. and {Titus}, Michael and {Toma}, Kenji and {Torne}, Pablo and {Trent}, Tyler and {Trippe}, Sascha and {Tsuda}, Shuichiro and {van Bemmel}, Ilse and {van Langevelde}, Huib Jan and {van Rossum}, Daniel R. and {Wagner}, Jan and {Wardle}, John and {Weintroub}, Jonathan and {Wex}, Norbert and {Wharton}, Robert and {Wielgus}, Maciek and {Wong}, George N. and {Wu}, Qingwen and {Young}, Ken and {Young}, Andr{\'e} and {Younsi}, Ziri and {Yuan}, Feng and {Yuan}, Ye-Fei and {Zensus}, J. Anton and {Zhao}, Guangyao and {Zhao}, Shan-Shan and {Zhu}, Ziyan and {Algaba}, Juan-Carlos and {Allardi}, Alexander and {Amestica}, Rodrigo and {Anczarski}, Jadyn and {Bach}, Uwe and {Baganoff}, Frederick K. and {Beaudoin}, Christopher and {Benson}, Bradford A. and {Berthold}, Ryan and {Blanchard}, Jay M. and {Blundell}, Ray and {Bustamente}, Sandra and {Cappallo}, Roger and {Castillo-Dom{\'\i}nguez}, Edgar and {Chang}, Chih-Cheng and {Chang}, Shu-Hao and {Chang}, Song-Chu and {Chen}, Chung-Chen and {Chilson}, Ryan and {Chuter}, Tim C. and {C{\'o}rdova Rosado}, Rodrigo and {Coulson}, Iain M. and {Crawford}, Thomas M. and {Crowley}, Joseph and {David}, John and {Derome}, Mark and {Dexter}, Matthew and {Dornbusch}, Sven and {Dudevoir}, Kevin A. and {Dzib}, Sergio A. and {Eckart}, Andreas and {Eckert}, Chris and {Erickson}, Neal R. and {Everett}, Wendeline B. and {Faber}, Aaron and {Farah}, Joseph R. and {Fath}, Vernon and {Folkers}, Thomas W. and {Forbes}, David C. and {Freund}, Robert and {G{\'o}mez-Ruiz}, Arturo I. and {Gale}, David M. and {Gao}, Feng and {Geertsema}, Gertie and {Graham}, David A. and {Greer}, Christopher H. and {Grosslein}, Ronald and {Gueth}, Fr{\'e}d{\'e}ric and {Haggard}, Daryl and {Halverson}, Nils W. and {Han}, Chih-Chiang and {Han}, Kuo-Chang and {Hao}, Jinchi and {Hasegawa}, Yutaka and {Henning}, Jason W. and {Hern{\'a}ndez-G{\'o}mez}, Antonio and {Herrero-Illana}, Rub{\'e}n and {Heyminck}, Stefan and {Hirota}, Akihiko and {Hoge}, James and {Huang}, Yau-De and {Impellizzeri}, C.~M. Violette and {Jiang}, Homin and {Kamble}, Atish and {Keisler}, Ryan and {Kimura}, Kimihiro and {Kono}, Yusuke and {Kubo}, Derek and {Kuroda}, John and {Lacasse}, Richard and {Laing}, Robert A. and {Leitch}, Erik M. and {Li}, Chao-Te and {Lin}, Lupin C. -C. and {Liu}, Ching-Tang and {Liu}, Kuan-Yu and {Lu}, Li-Ming and {Marson}, Ralph G. and {Martin-Cocher}, Pierre L. and {Massingill}, Kyle D. and {Matulonis}, Callie and {McColl}, Martin P. and {McWhirter}, Stephen R. and {Messias}, Hugo and {Meyer-Zhao}, Zheng and {Michalik}, Daniel and {Monta{\~n}a}, Alfredo and {Montgomerie}, William and {Mora-Klein}, Matias and {Muders}, Dirk and {Nadolski}, Andrew and {Navarro}, Santiago and {Neilsen}, Joseph and {Nguyen}, Chi H. and {Nishioka}, Hiroaki and {Norton}, Timothy and {Nowak}, Michael A. and {Nystrom}, George and {Ogawa}, Hideo and {Oshiro}, Peter and {Oyama}, Tomoaki and {Parsons}, Harriet and {Paine}, Scott N. and {Pe{\~n}alver}, Juan and {Phillips}, Neil M. and {Poirier}, Michael and {Pradel}, Nicolas and {Primiani}, Rurik A. and {Raffin}, Philippe A. and {Rahlin}, Alexandra S. and {Reiland}, George and {Risacher}, Christopher and {Ruiz}, Ignacio and {S{\'a}ez-Mada{\'\i}n}, Alejandro F. and {Sassella}, Remi and {Schellart}, Pim and {Shaw}, Paul and {Silva}, Kevin M. and {Shiokawa}, Hotaka and {Smith}, David R. and {Snow}, William and {Souccar}, Kamal and {Sousa}, Don and {Sridharan}, T.~K. and {Srinivasan}, Ranjani and {Stahm}, William and {Stark}, Anthony A. and {Story}, Kyle and {Timmer}, Sjoerd T. and {Vertatschitsch}, Laura and {Walther}, Craig and {Wei}, Ta-Shun and {Whitehorn}, Nathan and {Whitney}, Alan R. and {Woody}, David P. and {Wouterloot}, Jan G.~A. and {Wright}, Melvin and {Yamaguchi}, Paul and {Yu}, Chen-Yu and {Zeballos}, Milagros and {Zhang}, Shuo and {Ziurys}, Lucy},
        title = "{First M87 Event Horizon Telescope Results. I. The Shadow of the Supermassive Black Hole}",
      journal = {\apjl},
         year = 2019,
        month = apr,
       volume = {875},
       number = {1},
          eid = {L1},
        pages = {L1},
          doi = {10.3847/2041-8213/ab0ec7},
archivePrefix = {arXiv},
       eprint = {1906.11238},
 primaryClass = {astro-ph.GA},
       adsurl = {https://ui.adsabs.harvard.edu/abs/2019ApJ...875L...1E}
}

@ARTICLE{Medling2015,
       author = {{Medling}, Anne M. and {U}, Vivian and {Max}, Claire E. and {Sanders}, David B. and {Armus}, Lee and {Holden}, Bradford and {Mieda}, Etsuko and {Wright}, Shelley A. and {Larkin}, James E.},
        title = "{Following Black Hole Scaling Relations through Gas-rich Mergers}",
      journal = {\apj},
         year = 2015,
        month = apr,
       volume = {803},
       number = {2},
          eid = {61},
        pages = {61},
          doi = {10.1088/0004-637X/803/2/61},
archivePrefix = {arXiv},
       eprint = {1502.06617},
 primaryClass = {astro-ph.GA},
       adsurl = {https://ui.adsabs.harvard.edu/abs/2015ApJ...803...61M}
}

@ARTICLE{1997ApJ...482L.139K,
       author = {{Kormendy}, John and {Bender}, Ralf and {Magorrian}, John and {Tremaine}, Scott and {Gebhardt}, Karl and {Richstone}, Douglas and {Dressler}, Alan and {Faber}, S.~M. and {Grillmair}, Carl and {Lauer}, Tod R.},
        title = "{Spectroscopic Evidence for a Supermassive Black Hole in NCG 4486B}",
      journal = {\apjl},
         year = 1997,
        month = jun,
       volume = {482},
       number = {2},
        pages = {L139-L142},
          doi = {10.1086/310720},
archivePrefix = {arXiv},
       eprint = {astro-ph/9703188},
 primaryClass = {astro-ph},
       adsurl = {https://ui.adsabs.harvard.edu/abs/1997ApJ...482L.139K}
}

@ARTICLE{2009ApJS..180..225H,
       author = {{Hinshaw}, G. and {Weiland}, J.~L. and {Hill}, R.~S. and {Odegard}, N. and {Larson}, D. and {Bennett}, C.~L. and {Dunkley}, J. and {Gold}, B. and {Greason}, M.~R. and {Jarosik}, N. and {Komatsu}, E. and {Nolta}, M.~R. and {Page}, L. and {Spergel}, D.~N. and {Wollack}, E. and {Halpern}, M. and {Kogut}, A. and {Limon}, M. and {Meyer}, S.~S. and {Tucker}, G.~S. and {Wright}, E.~L.},
        title = "{Five-Year Wilkinson Microwave Anisotropy Probe Observations: Data Processing, Sky Maps, and Basic Results}",
      journal = {\apjs},
         year = 2009,
        month = feb,
       volume = {180},
       number = {2},
        pages = {225-245},
          doi = {10.1088/0067-0049/180/2/225},
archivePrefix = {arXiv},
       eprint = {0803.0732},
 primaryClass = {astro-ph},
       adsurl = {https://ui.adsabs.harvard.edu/abs/2009ApJS..180..225H}
}

@ARTICLE{Ghez2008,
       author = {{Ghez}, A.~M. and {Salim}, S. and {Weinberg}, N.~N. and {Lu}, J.~R. and {Do}, T. and {Dunn}, J.~K. and {Matthews}, K. and {Morris}, M.~R. and {Yelda}, S. and {Becklin}, E.~E. and {Kremenek}, T. and {Milosavljevic}, M. and {Naiman}, J.},
        title = "{Measuring Distance and Properties of the Milky Way's Central Supermassive Black Hole with Stellar Orbits}",
      journal = {\apj},
         year = 2008,
        month = dec,
       volume = {689},
       number = {2},
        pages = {1044-1062},
          doi = {10.1086/592738},
archivePrefix = {arXiv},
       eprint = {0808.2870},
 primaryClass = {astro-ph},
       adsurl = {https://ui.adsabs.harvard.edu/abs/2008ApJ...689.1044G}
}

@ARTICLE{Genzel2010,
       author = {{Genzel}, Reinhard and {Eisenhauer}, Frank and {Gillessen}, Stefan},
        title = "{The Galactic Center massive black hole and nuclear star cluster}",
      journal = {Reviews of Modern Physics},
         year = 2010,
        month = oct,
       volume = {82},
       number = {4},
        pages = {3121-3195},
          doi = {10.1103/RevModPhys.82.3121},
archivePrefix = {arXiv},
       eprint = {1006.0064},
 primaryClass = {astro-ph.GA},
       adsurl = {https://ui.adsabs.harvard.edu/abs/2010RvMP...82.3121G}
}

@ARTICLE{Medling2014,
       author = {{Medling}, Anne M. and {U}, Vivian and {Guedes}, Javiera and {Max}, Claire E. and {Mayer}, Lucio and {Armus}, Lee and {Holden}, Bradford and {Ro{\v{s}}kar}, Rok and {Sanders}, David},
        title = "{Stellar and Gaseous Nuclear Disks Observed in Nearby (U)LIRGs}",
      journal = {\apj},
         year = 2014,
        month = mar,
       volume = {784},
       number = {1},
          eid = {70},
        pages = {70},
          doi = {10.1088/0004-637X/784/1/70},
archivePrefix = {arXiv},
       eprint = {1401.7338},
 primaryClass = {astro-ph.GA},
       adsurl = {https://ui.adsabs.harvard.edu/abs/2014ApJ...784...70M}
}

@ARTICLE{Kormendy2013,
       author = {{Kormendy}, John and {Ho}, Luis C.},
        title = "{Coevolution (Or Not) of Supermassive Black Holes and Host Galaxies}",
      journal = {\araa},
         year = 2013,
        month = aug,
       volume = {51},
       number = {1},
        pages = {511-653},
          doi = {10.1146/annurev-astro-082708-101811},
archivePrefix = {arXiv},
       eprint = {1304.7762},
 primaryClass = {astro-ph.CO},
       adsurl = {https://ui.adsabs.harvard.edu/abs/2013ARA&A..51..511K}
}

@ARTICLE{GATOS2021,
       author = {{Garc{\'\i}a-Burillo}, S. and {Alonso-Herrero}, A. and {Ramos Almeida}, C. and {Gonz{\'a}lez-Mart{\'\i}n}, O. and {Combes}, F. and {Usero}, A. and {H{\"o}nig}, S. and {Querejeta}, M. and {Hicks}, E.~K.~S. and {Hunt}, L.~K. and {Rosario}, D. and {Davies}, R. and {Boorman}, P.~G. and {Bunker}, A.~J. and {Burtscher}, L. and {Colina}, L. and {D{\'\i}az-Santos}, T. and {Gandhi}, P. and {Garc{\'\i}a-Bernete}, I. and {Garc{\'\i}a-Lorenzo}, B. and {Ichikawa}, K. and {Imanishi}, M. and {Izumi}, T. and {Labiano}, A. and {Levenson}, N.~A. and {L{\'o}pez-Rodr{\'\i}guez}, E. and {Packham}, C. and {Pereira-Santaella}, M. and {Ricci}, C. and {Rigopoulou}, D. and {Rouan}, D. and {Shimizu}, T. and {Stalevski}, M. and {Wada}, K. and {Williamson}, D.},
        title = "{The Galaxy Activity, Torus, and Outflow Survey (GATOS). I. ALMA images of dusty molecular tori in Seyfert galaxies}",
      journal = {\aap},
         year = 2021,
        month = aug,
       volume = {652},
          eid = {A98},
        pages = {A98},
          doi = {10.1051/0004-6361/202141075},
archivePrefix = {arXiv},
       eprint = {2104.10227},
 primaryClass = {astro-ph.GA},
       adsurl = {https://ui.adsabs.harvard.edu/abs/2021A&A...652A..98G}
}

@ARTICLE{Cen2012,
       author = {{Cen}, Renyue},
        title = "{Physics of Coevolution of Galaxies and Supermassive Black Holes}",
      journal = {\apj},
         year = 2012,
        month = aug,
       volume = {755},
       number = {1},
          eid = {28},
        pages = {28},
          doi = {10.1088/0004-637X/755/1/28},
archivePrefix = {arXiv},
       eprint = {1102.0262},
 primaryClass = {astro-ph.CO},
       adsurl = {https://ui.adsabs.harvard.edu/abs/2012ApJ...755...28C}
}

@ARTICLE{Hopkins2012,
       author = {{Hopkins}, Philip F.},
        title = "{Dynamical delays between starburst and AGN activity in galaxy nuclei}",
      journal = {\mnras},
         year = 2012,
        month = feb,
       volume = {420},
       number = {1},
        pages = {L8-L12},
          doi = {10.1111/j.1745-3933.2011.01179.x},
archivePrefix = {arXiv},
       eprint = {1101.4230},
 primaryClass = {astro-ph.CO},
       adsurl = {https://ui.adsabs.harvard.edu/abs/2012MNRAS.420L...8H}
}

@ARTICLE{Medling2019,
       author = {{Medling}, Anne M. and {Privon}, George C. and {Barcos-Mu{\~n}oz}, Loreto and {Treister}, Ezequiel and {Cicone}, Claudia and {Messias}, Hugo and {Sanders}, David B. and {Scoville}, Nick and {U}, Vivian and {Armus}, Lee and {Bauer}, Franz E. and {Chang}, Chin-Shin and {Comerford}, Julia M. and {Evans}, Aaron S. and {Max}, Claire E. and {M{\"u}ller-S{\'a}nchez}, Francisco and {Nagar}, Neil and {Sheth}, Kartik},
        title = "{How to Fuel an AGN: Mapping Circumnuclear Gas in NGC 6240 with ALMA}",
      journal = {\apjl},
         year = 2019,
        month = nov,
       volume = {885},
       number = {1},
          eid = {L21},
        pages = {L21},
          doi = {10.3847/2041-8213/ab4db7},
archivePrefix = {arXiv},
       eprint = {1910.12967},
 primaryClass = {astro-ph.GA},
       adsurl = {https://ui.adsabs.harvard.edu/abs/2019ApJ...885L..21M}
}

@ARTICLE{2019ApJ...871..166U,
       author = {{U}, Vivian and {Medling}, Anne M. and {Inami}, Hanae and {Armus}, Lee and {D{\'\i}az-Santos}, Tanio and {Charmandaris}, Vassilis and {Howell}, Justin and {Stierwalt}, Sabrina and {Privon}, George C. and {Linden}, Sean T. and {Sanders}, David B. and {Max}, Claire E. and {Evans}, Aaron S. and {Barcos-Mu{\~n}oz}, Loreto and {Chiang}, Charleston W.~K. and {Appleton}, Phil and {Canalizo}, Gabriela and {Fazio}, Giovanni and {Iwasawa}, Kazushi and {Larson}, Kirsten and {Mazzarella}, Joseph and {Murphy}, Eric and {Rich}, Jeffrey and {Surace}, Jason},
        title = "{Keck OSIRIS AO LIRG Analysis (KOALA): Feedback in the Nuclei of Luminous Infrared Galaxies}",
      journal = {\apj},
         year = 2019,
        month = feb,
       volume = {871},
       number = {2},
          eid = {166},
        pages = {166},
          doi = {10.3847/1538-4357/aaf1c2},
       adsurl = {https://ui.adsabs.harvard.edu/abs/2019ApJ...871..166U}
}

@ARTICLE{Stierwalt2013,
       author = {{Stierwalt}, S. and {Armus}, L. and {Surace}, J.~A. and {Inami}, H. and {Petric}, A.~O. and {Diaz-Santos}, T. and {Haan}, S. and {Charmandaris}, V. and {Howell}, J. and {Kim}, D.~C. and {Marshall}, J. and {Mazzarella}, J.~M. and {Spoon}, H.~W.~W. and {Veilleux}, S. and {Evans}, A. and {Sanders}, D.~B. and {Appleton}, P. and {Bothun}, G. and {Bridge}, C.~R. and {Chan}, B. and {Frayer}, D. and {Iwasawa}, K. and {Kewley}, L.~J. and {Lord}, S. and {Madore}, B.~F. and {Melbourne}, J.~E. and {Murphy}, E.~J. and {Rich}, J.~A. and {Schulz}, B. and {Sturm}, E. and {Vavilkin}, T. and {Xu}, K.},
        title = "{Mid-infrared Properties of Nearby Luminous Infrared Galaxies. I. Spitzer Infrared Spectrograph Spectra for the GOALS Sample}",
      journal = {\apjs},
         year = 2013,
        month = may,
       volume = {206},
       number = {1},
          eid = {1},
        pages = {1},
          doi = {10.1088/0067-0049/206/1/1},
archivePrefix = {arXiv},
       eprint = {1302.4477},
 primaryClass = {astro-ph.CO},
       adsurl = {https://ui.adsabs.harvard.edu/abs/2013ApJS..206....1S}
}

@ARTICLE{2017ApJ...843..117B,
       author = {{Barcos-Mu{\~n}oz}, L. and {Leroy}, A.~K. and {Evans}, A.~S. and {Condon}, J. and {Privon}, G.~C. and {Thompson}, T.~A. and {Armus}, L. and {D{\'\i}az-Santos}, T. and {Mazzarella}, J.~M. and {Meier}, D.~S. and {Momjian}, E. and {Murphy}, E.~J. and {Ott}, J. and {Sanders}, D.~B. and {Schinnerer}, E. and {Stierwalt}, S. and {Surace}, J.~A. and {Walter}, F.},
        title = "{A 33 GHz Survey of Local Major Mergers: Estimating the Sizes of the Energetically Dominant Regions from High-resolution Measurements of the Radio Continuum}",
      journal = {\apj},
         year = 2017,
        month = jul,
       volume = {843},
       number = {2},
          eid = {117},
        pages = {117},
          doi = {10.3847/1538-4357/aa789a},
archivePrefix = {arXiv},
       eprint = {1705.10801},
 primaryClass = {astro-ph.GA},
       adsurl = {https://ui.adsabs.harvard.edu/abs/2017ApJ...843..117B}
}

@ARTICLE{2006ApJ...641..689V,
       author = {{Vestergaard}, Marianne and {Peterson}, Bradley M.},
        title = "{Determining Central Black Hole Masses in Distant Active Galaxies and Quasars. II. Improved Optical and UV Scaling Relationships}",
      journal = {\apj},
         year = 2006,
        month = apr,
       volume = {641},
       number = {2},
        pages = {689-709},
          doi = {10.1086/500572},
archivePrefix = {arXiv},
       eprint = {astro-ph/0601303},
 primaryClass = {astro-ph},
       adsurl = {https://ui.adsabs.harvard.edu/abs/2006ApJ...641..689V}
}

@ARTICLE{2004PASP..116..465H,
       author = {{Horne}, Keith and {Peterson}, Bradley M. and {Collier}, Stefan J. and {Netzer}, Hagai},
        title = "{Observational Requirements for High-Fidelity Reverberation Mapping}",
      journal = {\pasp},
         year = 2004,
        month = may,
       volume = {116},
       number = {819},
        pages = {465-476},
          doi = {10.1086/420755},
       adsurl = {https://ui.adsabs.harvard.edu/abs/2004PASP..116..465H}
}

@ARTICLE{1991ApJ...366...64C,
       author = {{Clavel}, J. and {Reichert}, G.~A. and {Alloin}, D. and {Crenshaw}, D.~M. and {Kriss}, G. and {Krolik}, J.~H. and {Malkan}, M.~A. and {Netzer}, H. and {Peterson}, B.~M. and {Wamsteker}, W. and {Altamore}, A. and {Baribaud}, T. and {Barr}, P. and {Beck}, S. and {Binette}, L. and {Bromage}, G.~E. and {Brosch}, N. and {Diaz}, A.~I. and {Filippenko}, A.~V. and {Fricke}, K. and {Gaskell}, C.~M. and {Giommi}, P. and {Glass}, I.~S. and {Gondhalekar}, P. and {Hackney}, R.~L. and {Halpern}, J.~P. and {Hutter}, D.~J. and {Joersaeter}, S. and {Kinney}, A.~L. and {Kollatschny}, W. and {Koratkar}, A. and {Korista}, K.~T. and {Laor}, A. and {Lasota}, J. -P. and {Leibowitz}, E. and {Maoz}, D. and {Martin}, P.~G. and {Mazeh}, T. and {Meurs}, E.~J.~A. and {Nair}, A.~D. and {O'Brien}, P. and {Pelat}, D. and {Perez}, E. and {Perola}, G.~C. and {Ptak}, R.~L. and {Rodriguez-Pascual}, P. and {Rosenblatt}, E.~I. and {Sadun}, A.~C. and {Santos-Lleo}, M. and {Shaw}, R.~A. and {Smith}, P.~S. and {Stirpe}, G.~M. and {Stoner}, R. and {Sun}, W.~H. and {Ulrich}, M. -H. and {van Groningen}, E. and {Zheng}, W.},
        title = "{Steps toward Determination of the Size and Structure of the Broad-Line Region in Active Galactic Nuclei. I. an 8 Month Campaign of Monitoring NGC 5548 with IUE}",
      journal = {\apj},
         year = 1991,
        month = jan,
       volume = {366},
        pages = {64},
          doi = {10.1086/169540},
       adsurl = {https://ui.adsabs.harvard.edu/abs/1991ApJ...366...64C}
}

@ARTICLE{2021iSci...24j2557C,
       author = {{Cackett}, Edward M. and {Bentz}, Misty C. and {Kara}, Erin},
        title = "{Reverberation mapping of active galactic nuclei: from X-ray corona to dusty torus}",
      journal = {iScience},
         year = 2021,
        month = jun,
       volume = {24},
       number = {6},
        pages = {102557},
          doi = {10.1016/j.isci.2021.102557},
archivePrefix = {arXiv},
       eprint = {2105.06926},
 primaryClass = {astro-ph.GA},
       adsurl = {https://ui.adsabs.harvard.edu/abs/2021iSci...24j2557C}
}

@ARTICLE{Scoville2016,
       author = {{Scoville}, N. and {Sheth}, K. and {Aussel}, H. and {Vanden Bout}, P. and {Capak}, P. and {Bongiorno}, A. and {Casey}, C.~M. and {Murchikova}, L. and {Koda}, J. and {{\'A}lvarez-M{\'a}rquez}, J. and {Lee}, N. and {Laigle}, C. and {McCracken}, H.~J. and {Ilbert}, O. and {Pope}, A. and {Sanders}, D. and {Chu}, J. and {Toft}, S. and {Ivison}, R.~J. and {Manohar}, S.},
        title = "{ISM Masses and the Star formation Law at Z = 1 to 6: ALMA Observations of Dust Continuum in 145 Galaxies in the COSMOS Survey Field}",
      journal = {\apj},
         year = 2016,
        month = apr,
       volume = {820},
       number = {2},
          eid = {83},
        pages = {83},
          doi = {10.3847/0004-637X/820/2/83},
archivePrefix = {arXiv},
       eprint = {1511.05149},
 primaryClass = {astro-ph.GA},
       adsurl = {https://ui.adsabs.harvard.edu/abs/2016ApJ...820...83S}
}

@ARTICLE{Faber1976,
       author = {{Faber}, S.~M. and {Jackson}, R.~E.},
        title = "{Velocity dispersions and mass-to-light ratios for elliptical galaxies.}",
      journal = {\apj},
         year = 1976,
        month = mar,
       volume = {204},
        pages = {668-683},
          doi = {10.1086/154215},
       adsurl = {https://ui.adsabs.harvard.edu/abs/1976ApJ...204..668F}
}

@ARTICLE{Lauer1995,
       author = {{Lauer}, T.~R. and {Ajhar}, E.~A. and {Byun}, Y. -I. and {Dressler}, A. and {Faber}, S.~M. and {Grillmair}, C. and {Kormendy}, J. and {Richstone}, D. and {Tremaine}, S.},
        title = "{The Centers of Early-Type Galaxies with HST.I.An Observational Survey}",
      journal = {\aj},
         year = 1995,
        month = dec,
       volume = {110},
        pages = {2622},
          doi = {10.1086/117719},
       adsurl = {https://ui.adsabs.harvard.edu/abs/1995AJ....110.2622L}
}

@ARTICLE{1999ApJS..124..383C,
       author = {{Cretton}, N. and {de Zeeuw}, P. Tim and {van der Marel}, Roeland P. and {Rix}, Hans-Walter},
        title = "{Axisymmetric Three-Integral Models for Galaxies}",
      journal = {\apjs},
         year = 1999,
        month = oct,
       volume = {124},
       number = {2},
        pages = {383-401},
          doi = {10.1086/313264},
archivePrefix = {arXiv},
       eprint = {astro-ph/9902034},
 primaryClass = {astro-ph},
       adsurl = {https://ui.adsabs.harvard.edu/abs/1999ApJS..124..383C}
}

@ARTICLE{Gebhardt2000,
       author = {{Gebhardt}, Karl and {Bender}, Ralf and {Bower}, Gary and {Dressler}, Alan and {Faber}, S.~M. and {Filippenko}, Alexei V. and {Green}, Richard and {Grillmair}, Carl and {Ho}, Luis C. and {Kormendy}, John and {Lauer}, Tod R. and {Magorrian}, John and {Pinkney}, Jason and {Richstone}, Douglas and {Tremaine}, Scott},
        title = "{A Relationship between Nuclear Black Hole Mass and Galaxy Velocity Dispersion}",
      journal = {\apjl},
         year = 2000,
        month = aug,
       volume = {539},
       number = {1},
        pages = {L13-L16},
          doi = {10.1086/312840},
archivePrefix = {arXiv},
       eprint = {astro-ph/0006289},
 primaryClass = {astro-ph},
       adsurl = {https://ui.adsabs.harvard.edu/abs/2000ApJ...539L..13G}
}

@ARTICLE{vandenbosch2008,
       author = {{van den Bosch}, R.~C.~E. and {van de Ven}, G. and {Verolme}, E.~K. and {Cappellari}, M. and {de Zeeuw}, P.~T.},
        title = "{Triaxial orbit based galaxy models with an application to the (apparent) decoupled core galaxy NGC 4365}",
      journal = {\mnras},
         year = 2008,
        month = apr,
       volume = {385},
       number = {2},
        pages = {647-666},
          doi = {10.1111/j.1365-2966.2008.12874.x},
archivePrefix = {arXiv},
       eprint = {0712.0113},
 primaryClass = {astro-ph},
       adsurl = {https://ui.adsabs.harvard.edu/abs/2008MNRAS.385..647V}
}

@ARTICLE{Walsh2012,
       author = {{Walsh}, Jonelle L. and {van den Bosch}, Remco C.~E. and {Barth}, Aaron J. and {Sarzi}, Marc},
        title = "{A Stellar Dynamical Mass Measurement of the Black Hole in NGC 3998 from Keck Adaptive Optics Observations}",
      journal = {\apj},
         year = 2012,
        month = jul,
       volume = {753},
       number = {1},
          eid = {79},
        pages = {79},
          doi = {10.1088/0004-637X/753/1/79},
archivePrefix = {arXiv},
       eprint = {1205.0816},
 primaryClass = {astro-ph.CO},
       adsurl = {https://ui.adsabs.harvard.edu/abs/2012ApJ...753...79W}
}

@ARTICLE{vestergaard2006,
       author = {{Vestergaard}, Marianne and {Peterson}, Bradley M.},
        title = "{Determining Central Black Hole Masses in Distant Active Galaxies and Quasars. II. Improved Optical and UV Scaling Relationships}",
      journal = {\apj},
         year = 2006,
        month = apr,
       volume = {641},
       number = {2},
        pages = {689-709},
          doi = {10.1086/500572},
archivePrefix = {arXiv},
       eprint = {astro-ph/0601303},
 primaryClass = {astro-ph},
       adsurl = {https://ui.adsabs.harvard.edu/abs/2006ApJ...641..689V}
}

@ARTICLE{Davis2018,
       author = {{Davis}, Benjamin L. and {Graham}, Alister W. and {Cameron}, Ewan},
        title = "{Black Hole Mass Scaling Relations for Spiral Galaxies. II. M $_{BH}$-M $_{*,tot}$ and M $_{BH}$-M $_{*,disk}$}",
      journal = {\apj},
         year = 2018,
        month = dec,
       volume = {869},
       number = {2},
          eid = {113},
        pages = {113},
          doi = {10.3847/1538-4357/aae820},
archivePrefix = {arXiv},
       eprint = {1810.04888},
 primaryClass = {astro-ph.GA},
       adsurl = {https://ui.adsabs.harvard.edu/abs/2018ApJ...869..113D}
}

@ARTICLE{Davis2019,
       author = {{Davis}, Benjamin L. and {Graham}, Alister W. and {Cameron}, Ewan},
        title = "{Black Hole Mass Scaling Relations for Spiral Galaxies. I. M $_{BH}$-M $_{*,sph}$}",
      journal = {\apj},
         year = 2019,
        month = mar,
       volume = {873},
       number = {1},
          eid = {85},
        pages = {85},
          doi = {10.3847/1538-4357/aaf3b8},
archivePrefix = {arXiv},
       eprint = {1810.04887},
 primaryClass = {astro-ph.GA},
       adsurl = {https://ui.adsabs.harvard.edu/abs/2019ApJ...873...85D}
}

@ARTICLE{2009PASP..121..559A,
       author = {{Armus}, L. and {Mazzarella}, J.~M. and {Evans}, A.~S. and {Surace}, J.~A. and {Sanders}, D.~B. and {Iwasawa}, K. and {Frayer}, D.~T. and {Howell}, J.~H. and {Chan}, B. and {Petric}, A. and {Vavilkin}, T. and {Kim}, D.~C. and {Haan}, S. and {Inami}, H. and {Murphy}, E.~J. and {Appleton}, P.~N. and {Barnes}, J.~E. and {Bothun}, G. and {Bridge}, C.~R. and {Charmandaris}, V. and {Jensen}, J.~B. and {Kewley}, L.~J. and {Lord}, S. and {Madore}, B.~F. and {Marshall}, J.~A. and {Melbourne}, J.~E. and {Rich}, J. and {Satyapal}, S. and {Schulz}, B. and {Spoon}, H.~W.~W. and {Sturm}, E. and {U}, V. and {Veilleux}, S. and {Xu}, K.},
        title = "{GOALS: The Great Observatories All-Sky LIRG Survey}",
      journal = {\pasp},
         year = 2009,
        month = jun,
       volume = {121},
       number = {880},
        pages = {559},
          doi = {10.1086/600092},
archivePrefix = {arXiv},
       eprint = {0904.4498},
 primaryClass = {astro-ph.CO},
       adsurl = {https://ui.adsabs.harvard.edu/abs/2009PASP..121..559A}
}

@ARTICLE{Mirabel1988,
       author = {{Mirabel}, I.~F. and {Sanders}, D.~B.},
        title = "{21 Centimeter Survey of Luminous Infrared Galaxies}",
      journal = {\apj},
         year = 1988,
        month = dec,
       volume = {335},
        pages = {104},
          doi = {10.1086/166909},
       adsurl = {https://ui.adsabs.harvard.edu/abs/1988ApJ...335..104M}
}

@ARTICLE{Hopkins2008a,
       author = {{Hopkins}, Philip F. and {Hernquist}, Lars and {Cox}, Thomas J. and {Kere{\v{s}}}, Du{\v{s}}an},
        title = "{A Cosmological Framework for the Co-Evolution of Quasars, Supermassive Black Holes, and Elliptical Galaxies. I. Galaxy Mergers and Quasar Activity}",
      journal = {\apjs},
         year = 2008,
        month = apr,
       volume = {175},
       number = {2},
        pages = {356-389},
          doi = {10.1086/524362},
archivePrefix = {arXiv},
       eprint = {0706.1243},
 primaryClass = {astro-ph},
       adsurl = {https://ui.adsabs.harvard.edu/abs/2008ApJS..175..356H}
}

@ARTICLE{Hopkins2008b,
       author = {{Hopkins}, Philip F. and {Cox}, Thomas J. and {Kere{\v{s}}}, Du{\v{s}}an and {Hernquist}, Lars},
        title = "{A Cosmological Framework for the Co-Evolution of Quasars, Supermassive Black Holes, and Elliptical Galaxies. II. Formation of Red Ellipticals}",
      journal = {\apjs},
         year = 2008,
        month = apr,
       volume = {175},
       number = {2},
        pages = {390-422},
          doi = {10.1086/524363},
archivePrefix = {arXiv},
       eprint = {0706.1246},
 primaryClass = {astro-ph},
       adsurl = {https://ui.adsabs.harvard.edu/abs/2008ApJS..175..390H}
}

@ARTICLE{Solomon2005,
       author = {{Solomon}, P.~M. and {Vanden Bout}, P.~A.},
        title = "{Molecular Gas at High Redshift}",
      journal = {\araa},
         year = 2005,
        month = sep,
       volume = {43},
       number = {1},
        pages = {677-725},
          doi = {10.1146/annurev.astro.43.051804.102221},
archivePrefix = {arXiv},
       eprint = {astro-ph/0508481},
 primaryClass = {astro-ph},
       adsurl = {https://ui.adsabs.harvard.edu/abs/2005ARA&A..43..677S}
}

@ARTICLE{Cicone2018,
       author = {{Cicone}, Claudia and {Severgnini}, Paola and {Papadopoulos}, Padelis P. and {Maiolino}, Roberto and {Feruglio}, Chiara and {Treister}, Ezequiel and {Privon}, George C. and {Zhang}, Zhi-yu and {Della Ceca}, Roberto and {Fiore}, Fabrizio and {Schawinski}, Kevin and {Wagg}, Jeff},
        title = "{ALMA [C I]$^{3}$ P $_{1}$-$^{3}$ P $_{0}$ Observations of NGC 6240: A Puzzling Molecular Outflow, and the Role of Outflows in the Global {\ensuremath{\alpha}} $_{CO}$ Factor of (U)LIRGs}",
      journal = {\apj},
         year = 2018,
        month = aug,
       volume = {863},
       number = {2},
          eid = {143},
        pages = {143},
          doi = {10.3847/1538-4357/aad32a},
archivePrefix = {arXiv},
       eprint = {1807.06015},
 primaryClass = {astro-ph.GA},
       adsurl = {https://ui.adsabs.harvard.edu/abs/2018ApJ...863..143C}
}

@ARTICLE{Sakamoto2021,
       author = {{Sakamoto}, Kazushi and {Mart{\'\i}n}, Sergio and {Wilner}, David J. and {Aalto}, Susanne and {Evans}, Aaron S. and {Harada}, Nanase},
        title = "{Deeply Buried Nuclei in the Infrared-luminous Galaxies NGC 4418 and Arp 220. II. Line Forests at {\ensuremath{\lambda}} = 1.4-0.4 mm and Circumnuclear Gas Observed with ALMA}",
      journal = {\apj},
         year = 2021,
        month = dec,
       volume = {923},
       number = {2},
          eid = {240},
        pages = {240},
          doi = {10.3847/1538-4357/ac29bf},
archivePrefix = {arXiv},
       eprint = {2109.08437},
 primaryClass = {astro-ph.GA},
       adsurl = {https://ui.adsabs.harvard.edu/abs/2021ApJ...923..240S}
}

@ARTICLE{2022ApJ...927...21W,
       author = {{Walter}, Fabian and {Neeleman}, Marcel and {Decarli}, Roberto and {Venemans}, Bram and {Meyer}, Romain and {Weiss}, Axel and {Ba{\~n}ados}, Eduardo and {Bosman}, Sarah E.~I. and {Carilli}, Chris and {Fan}, Xiaohui and {Riechers}, Dominik and {Rix}, Hans-Walter and {Thompson}, Todd A.},
        title = "{ALMA 200 pc Imaging of a z   7 Quasar Reveals a Compact, Disk-like Host Galaxy}",
      journal = {\apj},
         year = 2022,
        month = mar,
       volume = {927},
       number = {1},
          eid = {21},
        pages = {21},
          doi = {10.3847/1538-4357/ac49e8},
archivePrefix = {arXiv},
       eprint = {2201.06396},
 primaryClass = {astro-ph.GA},
       adsurl = {https://ui.adsabs.harvard.edu/abs/2022ApJ...927...21W}
}

@BOOK{Condon2016,
       author = {{Condon}, James J. and {Ransom}, Scott M.},
        title = "{Essential Radio Astronomy}",
         year = 2016,
       adsurl = {https://ui.adsabs.harvard.edu/abs/2016era..book.....C}
}

@ARTICLE{Scoville2015,
       author = {{Scoville}, Nick and {Sheth}, Kartik and {Walter}, Fabian and {Manohar}, Swarnima and {Zschaechner}, Laura and {Yun}, Min and {Koda}, Jin and {Sanders}, David and {Murchikova}, Lena and {Thompson}, Todd and {Robertson}, Brant and {Genzel}, Reinhard and {Hernquist}, Lars and {Tacconi}, Linda and {Brown}, Robert and {Narayanan}, Desika and {Hayward}, Christopher C. and {Barnes}, Joshua and {Kartaltepe}, Jeyhan and {Davies}, Richard and {van der Werf}, Paul and {Fomalont}, Edward},
        title = "{ALMA Imaging of HCN, CS, and Dust in Arp 220 and NGC 6240}",
      journal = {\apj},
         year = 2015,
        month = feb,
       volume = {800},
       number = {1},
          eid = {70},
        pages = {70},
          doi = {10.1088/0004-637X/800/1/70},
archivePrefix = {arXiv},
       eprint = {1412.5183},
 primaryClass = {astro-ph.GA},
       adsurl = {https://ui.adsabs.harvard.edu/abs/2015ApJ...800...70S}
}

@ARTICLE{2008MNRAS.390...71C,
       author = {{Cappellari}, Michele},
        title = "{Measuring the inclination and mass-to-light ratio of axisymmetric galaxies via anisotropic Jeans models of stellar kinematics}",
      journal = {\mnras},
         year = 2008,
        month = oct,
       volume = {390},
       number = {1},
        pages = {71-86},
          doi = {10.1111/j.1365-2966.2008.13754.x},
archivePrefix = {arXiv},
       eprint = {0806.0042},
 primaryClass = {astro-ph},
       adsurl = {https://ui.adsabs.harvard.edu/abs/2008MNRAS.390...71C}
}

@ARTICLE{Medling2011,
       author = {{Medling}, Anne M. and {Ammons}, S. Mark and {Max}, Claire E. and {Davies}, Richard I. and {Engel}, Hauke and {Canalizo}, Gabriela},
        title = "{Mass of the Southern Black Hole in NGC 6240 from Laser Guide Star Adaptive Optics}",
      journal = {\apj},
         year = 2011,
        month = dec,
       volume = {743},
       number = {1},
          eid = {32},
        pages = {32},
          doi = {10.1088/0004-637X/743/1/32},
archivePrefix = {arXiv},
       eprint = {1108.5180},
 primaryClass = {astro-ph.CO},
       adsurl = {https://ui.adsabs.harvard.edu/abs/2011ApJ...743...32M}
}

@ARTICLE{Dunne2011,
       author = {{Dunne}, L. and {Gomez}, H.~L. and {da Cunha}, E. and {Charlot}, S. and {Dye}, S. and {Eales}, S. and {Maddox}, S.~J. and {Rowlands}, K. and {Smith}, D.~J.~B. and {Auld}, R. and {Baes}, M. and {Bonfield}, D.~G. and {Bourne}, N. and {Buttiglione}, S. and {Cava}, A. and {Clements}, D.~L. and {Coppin}, K.~E.~K. and {Cooray}, A. and {Dariush}, A. and {de Zotti}, G. and {Driver}, S. and {Fritz}, J. and {Geach}, J. and {Hopwood}, R. and {Ibar}, E. and {Ivison}, R.~J. and {Jarvis}, M.~J. and {Kelvin}, L. and {Pascale}, E. and {Pohlen}, M. and {Popescu}, C. and {Rigby}, E.~E. and {Robotham}, A. and {Rodighiero}, G. and {Sansom}, A.~E. and {Serjeant}, S. and {Temi}, P. and {Thompson}, M. and {Tuffs}, R. and {van der Werf}, P. and {Vlahakis}, C.},
        title = "{Herschel-ATLAS: rapid evolution of dust in galaxies over the last 5 billion years}",
      journal = {\mnras},
         year = 2011,
        month = oct,
       volume = {417},
       number = {2},
        pages = {1510-1533},
          doi = {10.1111/j.1365-2966.2011.19363.x},
archivePrefix = {arXiv},
       eprint = {1012.5186},
 primaryClass = {astro-ph.CO},
       adsurl = {https://ui.adsabs.harvard.edu/abs/2011MNRAS.417.1510D}
}

@ARTICLE{Dale2012,
       author = {{Dale}, D.~A. and {Aniano}, G. and {Engelbracht}, C.~W. and {Hinz}, J.~L. and {Krause}, O. and {Montiel}, E.~J. and {Roussel}, H. and {Appleton}, P.~N. and {Armus}, L. and {Beir{\~a}o}, P. and {Bolatto}, A.~D. and {Brandl}, B.~R. and {Calzetti}, D. and {Crocker}, A.~F. and {Croxall}, K.~V. and {Draine}, B.~T. and {Galametz}, M. and {Gordon}, K.~D. and {Groves}, B.~A. and {Hao}, C. -N. and {Helou}, G. and {Hunt}, L.~K. and {Johnson}, B.~D. and {Kennicutt}, R.~C. and {Koda}, J. and {Leroy}, A.~K. and {Li}, Y. and {Meidt}, S.~E. and {Miller}, A.~E. and {Murphy}, E.~J. and {Rahman}, N. and {Rix}, H. -W. and {Sandstrom}, K.~M. and {Sauvage}, M. and {Schinnerer}, E. and {Skibba}, R.~A. and {Smith}, J. -D.~T. and {Tabatabaei}, F.~S. and {Walter}, F. and {Wilson}, C.~D. and {Wolfire}, M.~G. and {Zibetti}, S.},
        title = "{Herschel Far-infrared and Submillimeter Photometry for the KINGFISH Sample of nearby Galaxies}",
      journal = {\apj},
         year = 2012,
        month = jan,
       volume = {745},
       number = {1},
          eid = {95},
        pages = {95},
          doi = {10.1088/0004-637X/745/1/95},
archivePrefix = {arXiv},
       eprint = {1112.1093},
 primaryClass = {astro-ph.CO},
       adsurl = {https://ui.adsabs.harvard.edu/abs/2012ApJ...745...95D}
}

@ARTICLE{Auld2013,
       author = {{Auld}, R. and {Bianchi}, S. and {Smith}, M.~W.~L. and {Davies}, J.~I. and {Bendo}, G.~J. and {di Serego}, S. Alighieri and {Cortese}, L. and {Baes}, M. and {Bomans}, D.~J. and {Boquien}, M. and {Boselli}, A. and {Ciesla}, L. and {Clemens}, M. and {Corbelli}, E. and {De Looze}, I. and {Fritz}, J. and {Gavazzi}, G. and {Pappalardo}, C. and {Grossi}, M. and {Hunt}, L.~K. and {Madden}, S. and {Magrini}, L. and {Pohlen}, M. and {Verstappen}, J. and {Vlahakis}, C. and {Xilouris}, E.~M. and {Zibetti}, S.},
        title = "{The Herschel Virgo Cluster Survey - XII. FIR properties of optically selected Virgo cluster galaxies}",
      journal = {\mnras},
         year = 2013,
        month = jan,
       volume = {428},
       number = {3},
        pages = {1880-1910},
          doi = {10.1093/mnras/sts125},
archivePrefix = {arXiv},
       eprint = {1209.4651},
 primaryClass = {astro-ph.CO},
       adsurl = {https://ui.adsabs.harvard.edu/abs/2013MNRAS.428.1880A}
}

@ARTICLE{Bondi1952,
       author = {{Bondi}, H.},
        title = "{On spherically symmetrical accretion}",
      journal = {\mnras},
         year = 1952,
        month = jan,
       volume = {112},
        pages = {195},
          doi = {10.1093/mnras/112.2.195},
       adsurl = {https://ui.adsabs.harvard.edu/abs/1952MNRAS.112..195B}
}

@ARTICLE{Hoyle1939,
       author = {{Hoyle}, F. and {Lyttleton}, R.~A.},
        title = "{The effect of interstellar matter on climatic variation}",
      journal = {Proceedings of the Cambridge Philosophical Society},
         year = 1939,
        month = jan,
       volume = {35},
       number = {3},
        pages = {405},
          doi = {10.1017/S0305004100021150},
       adsurl = {https://ui.adsabs.harvard.edu/abs/1939PCPS...35..405H}
}

@ARTICLE{1979SvA....23..201B,
       author = {{Bisnovatyi-Kogan}, G.~S. and {Kazhdan}, Ya. M. and {Klypin}, A.~A. and {Lutskii}, A.~E. and {Shakura}, N.~I.},
        title = "{Accretion onto a rapidly moving gravitating center}",
      journal = {\sovast},
         year = 1979,
        month = apr,
       volume = {23},
        pages = {201-205},
       adsurl = {https://ui.adsabs.harvard.edu/abs/1979SvA....23..201B}
}

@ARTICLE{1944MNRAS.104..273B,
       author = {{Bondi}, H. and {Hoyle}, F.},
        title = "{On the mechanism of accretion by stars}",
      journal = {\mnras},
         year = 1944,
        month = jan,
       volume = {104},
        pages = {273},
          doi = {10.1093/mnras/104.5.273},
       adsurl = {https://ui.adsabs.harvard.edu/abs/1944MNRAS.104..273B}
}

@ARTICLE{Power2011,
       author = {{Power}, Chris and {Nayakshin}, Sergei and {King}, Andrew},
        title = "{The accretion disc particle method for simulations of black hole feeding and feedback}",
      journal = {\mnras},
         year = 2011,
        month = mar,
       volume = {412},
       number = {1},
        pages = {269-276},
          doi = {10.1111/j.1365-2966.2010.17901.x},
archivePrefix = {arXiv},
       eprint = {1003.0605},
 primaryClass = {astro-ph.CO},
       adsurl = {https://ui.adsabs.harvard.edu/abs/2011MNRAS.412..269P}
}

@ARTICLE{Mayer2007,
       author = {{Mayer}, L. and {Kazantzidis}, S. and {Madau}, P. and {Colpi}, M. and {Quinn}, T. and {Wadsley}, J.},
        title = "{Rapid Formation of Supermassive Black Hole Binaries in Galaxy Mergers with Gas}",
      journal = {Science},
         year = 2007,
        month = jun,
       volume = {316},
       number = {5833},
        pages = {1874},
          doi = {10.1126/science.1141858},
archivePrefix = {arXiv},
       eprint = {0706.1562},
 primaryClass = {astro-ph},
       adsurl = {https://ui.adsabs.harvard.edu/abs/2007Sci...316.1874M}
}

@ARTICLE{Matplotlib,
  author={Hunter, John D.},
  journal={Computing in Science \& Engineering}, 
  title={Matplotlib: A 2D Graphics Environment}, 
  year={2007},
  volume={9},
  number={3},
  pages={90-95},
  doi={10.1109/MCSE.2007.55}}

@ARTICLE{harris2020array,
 title         = {Array programming with {NumPy}},
 author        = {Charles R. Harris and K. Jarrod Millman and St{\'{e}}fan J.
                 van der Walt and Ralf Gommers and Pauli Virtanen and David
                 Cournapeau and Eric Wieser and Julian Taylor and Sebastian
                 Berg and Nathaniel J. Smith and Robert Kern and Matti Picus
                 and Stephan Hoyer and Marten H. van Kerkwijk and Matthew
                 Brett and Allan Haldane and Jaime Fern{\'{a}}ndez del
                 R{\'{i}}o and Mark Wiebe and Pearu Peterson and Pierre
                 G{\'{e}}rard-Marchant and Kevin Sheppard and Tyler Reddy and
                 Warren Weckesser and Hameer Abbasi and Christoph Gohlke and
                 Travis E. Oliphant},
 year          = {2020},
 month         = sep,
 journal       = {Nature},
 volume        = {585},
 number        = {7825},
 pages         = {357--362},
 doi           = {10.1038/s41586-020-2649-2},
 publisher     = {Springer Science and Business Media {LLC}},
 url           = {https://doi.org/10.1038/s41586-020-2649-2}
}

@INPROCEEDINGS{Kormendy1993a,
       author = {{Kormendy}, J.},
        title = "{A critical review of stellar-dynamical evidence for black holes in galaxy nuclei}",
    booktitle = {The Nearest Active Galaxies},
         year = 1993,
       editor = {{Beckman}, John and {Colina}, Luis and {Netzer}, Hagai},
        month = jan,
        pages = {197-218},
       adsurl = {https://ui.adsabs.harvard.edu/abs/1993nag..conf..197K}
}

@INPROCEEDINGS{Ho1999a,
       author = {{Ho}, Luis},
        title = "{Supermassive Black Holes in Galactic Nuclei: Observational Evidence and Astrophysical Consequences}",
    booktitle = {Observational Evidence for the Black Holes in the Universe},
         year = 1999,
       editor = {{Chakrabarti}, Sandip K.},
       series = {Astrophysics and Space Science Library},
       volume = {234},
        month = jan,
        pages = {157},
          doi = {10.1007/978-94-011-4750-7_11},
       adsurl = {https://ui.adsabs.harvard.edu/abs/1999ASSL..234..157H}
}

@ARTICLE{McConnellMa2013,
       author = {{McConnell}, Nicholas J. and {Ma}, Chung-Pei},
        title = "{Revisiting the Scaling Relations of Black Hole Masses and Host Galaxy Properties}",
      journal = {\apj},
         year = 2013,
        month = feb,
       volume = {764},
       number = {2},
          eid = {184},
        pages = {184},
          doi = {10.1088/0004-637X/764/2/184},
archivePrefix = {arXiv},
       eprint = {1211.2816},
 primaryClass = {astro-ph.CO},
       adsurl = {https://ui.adsabs.harvard.edu/abs/2013ApJ...764..184M}
}

@ARTICLE{MerrittFerrarese2001,
       author = {{Merritt}, David and {Ferrarese}, Laura},
        title = "{Black hole demographics from the M$_{{\textbullet}}$-{\ensuremath{\sigma}} relation}",
      journal = {\mnras},
         year = 2001,
        month = jan,
       volume = {320},
       number = {3},
        pages = {L30-L34},
          doi = {10.1046/j.1365-8711.2001.04165.x},
archivePrefix = {arXiv},
       eprint = {astro-ph/0009076},
 primaryClass = {astro-ph},
       adsurl = {https://ui.adsabs.harvard.edu/abs/2001MNRAS.320L..30M}
}

@INPROCEEDINGS{KormendyGebhart2001,
       author = {{Kormendy}, John and {Gebhardt}, Karl},
        title = "{Supermassive black holes in galactic nuclei}",
    booktitle = {20th Texas Symposium on relativistic astrophysics},
         year = 2001,
       editor = {{Wheeler}, J. Craig and {Martel}, Hugo},
       series = {American Institute of Physics Conference Series},
       volume = {586},
        month = oct,
        pages = {363-381},
          doi = {10.1063/1.1419581},
archivePrefix = {arXiv},
       eprint = {astro-ph/0105230},
 primaryClass = {astro-ph},
       adsurl = {https://ui.adsabs.harvard.edu/abs/2001AIPC..586..363K}
}

@ARTICLE{MclureDunlop2002,
       author = {{McLure}, R.~J. and {Dunlop}, J.~S.},
        title = "{On the black hole-bulge mass relation in active and inactive galaxies}",
      journal = {\mnras},
         year = 2002,
        month = apr,
       volume = {331},
       number = {3},
        pages = {795-804},
          doi = {10.1046/j.1365-8711.2002.05236.x},
archivePrefix = {arXiv},
       eprint = {astro-ph/0108417},
 primaryClass = {astro-ph},
       adsurl = {https://ui.adsabs.harvard.edu/abs/2002MNRAS.331..795M}
}

@ARTICLE{MarconiHunt2003,
       author = {{Marconi}, Alessandro and {Hunt}, Leslie K.},
        title = "{The Relation between Black Hole Mass, Bulge Mass, and Near-Infrared Luminosity}",
      journal = {\apjl},
         year = 2003,
        month = may,
       volume = {589},
       number = {1},
        pages = {L21-L24},
          doi = {10.1086/375804},
archivePrefix = {arXiv},
       eprint = {astro-ph/0304274},
 primaryClass = {astro-ph},
       adsurl = {https://ui.adsabs.harvard.edu/abs/2003ApJ...589L..21M}
}

@ARTICLE{HaringRix2004,
       author = {{H{\"a}ring}, Nadine and {Rix}, Hans-Walter},
        title = "{On the Black Hole Mass-Bulge Mass Relation}",
      journal = {\apjl},
         year = 2004,
        month = apr,
       volume = {604},
       number = {2},
        pages = {L89-L92},
          doi = {10.1086/383567},
archivePrefix = {arXiv},
       eprint = {astro-ph/0402376},
 primaryClass = {astro-ph},
       adsurl = {https://ui.adsabs.harvard.edu/abs/2004ApJ...604L..89H}
}

@ARTICLE{SilkRees1998,
       author = {{Silk}, Joseph and {Rees}, Martin J.},
        title = "{Quasars and galaxy formation}",
      journal = {\aap},
         year = 1998,
        month = mar,
       volume = {331},
        pages = {L1-L4},
          doi = {10.48550/arXiv.astro-ph/9801013},
archivePrefix = {arXiv},
       eprint = {astro-ph/9801013},
 primaryClass = {astro-ph},
       adsurl = {https://ui.adsabs.harvard.edu/abs/1998A&A...331L...1S}
}

@ARTICLE{Tremaine2002,
       author = {{Tremaine}, Scott and {Gebhardt}, Karl and {Bender}, Ralf and {Bower}, Gary and {Dressler}, Alan and {Faber}, S.~M. and {Filippenko}, Alexei V. and {Green}, Richard and {Grillmair}, Carl and {Ho}, Luis C. and {Kormendy}, John and {Lauer}, Tod R. and {Magorrian}, John and {Pinkney}, Jason and {Richstone}, Douglas},
        title = "{The Slope of the Black Hole Mass versus Velocity Dispersion Correlation}",
      journal = {\apj},
         year = 2002,
        month = aug,
       volume = {574},
       number = {2},
        pages = {740-753},
          doi = {10.1086/341002},
archivePrefix = {arXiv},
       eprint = {astro-ph/0203468},
 primaryClass = {astro-ph},
       adsurl = {https://ui.adsabs.harvard.edu/abs/2002ApJ...574..740T}
}

@ARTICLE{Gultekin2009c,
       author = {{G{\"u}ltekin}, Kayhan and {Richstone}, Douglas O. and {Gebhardt}, Karl and {Lauer}, Tod R. and {Tremaine}, Scott and {Aller}, M.~C. and {Bender}, Ralf and {Dressler}, Alan and {Faber}, S.~M. and {Filippenko}, Alexei V. and {Green}, Richard and {Ho}, Luis C. and {Kormendy}, John and {Magorrian}, John and {Pinkney}, Jason and {Siopis}, Christos},
        title = "{The M-{\ensuremath{\sigma}} and M-L Relations in Galactic Bulges, and Determinations of Their Intrinsic Scatter}",
      journal = {\apj},
         year = 2009,
        month = jun,
       volume = {698},
       number = {1},
        pages = {198-221},
          doi = {10.1088/0004-637X/698/1/198},
archivePrefix = {arXiv},
       eprint = {0903.4897},
 primaryClass = {astro-ph.GA},
       adsurl = {https://ui.adsabs.harvard.edu/abs/2009ApJ...698..198G}
}

@ARTICLE{Beifioiri2012,
       author = {{Beifiori}, A. and {Courteau}, S. and {Corsini}, E.~M. and {Zhu}, Y.},
        title = "{On the correlations between galaxy properties and supermassive black hole mass}",
      journal = {\mnras},
         year = 2012,
        month = jan,
       volume = {419},
       number = {3},
        pages = {2497-2528},
          doi = {10.1111/j.1365-2966.2011.19903.x},
archivePrefix = {arXiv},
       eprint = {1109.6265},
 primaryClass = {astro-ph.CO},
       adsurl = {https://ui.adsabs.harvard.edu/abs/2012MNRAS.419.2497B}
}

@ARTICLE{Bentz2013,
       author = {{Bentz}, Misty C. and {Denney}, Kelly D. and {Grier}, Catherine J. and {Barth}, Aaron J. and {Peterson}, Bradley M. and {Vestergaard}, Marianne and {Bennert}, Vardha N. and {Canalizo}, Gabriela and {De Rosa}, Gisella and {Filippenko}, Alexei V. and {Gates}, Elinor L. and {Greene}, Jenny E. and {Li}, Weidong and {Malkan}, Matthew A. and {Pogge}, Richard W. and {Stern}, Daniel and {Treu}, Tommaso and {Woo}, Jong-Hak},
        title = "{The Low-luminosity End of the Radius-Luminosity Relationship for Active Galactic Nuclei}",
      journal = {\apj},
         year = 2013,
        month = apr,
       volume = {767},
       number = {2},
          eid = {149},
        pages = {149},
          doi = {10.1088/0004-637X/767/2/149},
archivePrefix = {arXiv},
       eprint = {1303.1742},
 primaryClass = {astro-ph.CO},
       adsurl = {https://ui.adsabs.harvard.edu/abs/2013ApJ...767..149B}
}

@ARTICLE{Morgan2010,
       author = {{Morgan}, Christopher W. and {Kochanek}, C.~S. and {Morgan}, Nicholas D. and {Falco}, Emilio E.},
        title = "{The Quasar Accretion Disk Size-Black Hole Mass Relation}",
      journal = {\apj},
         year = 2010,
        month = apr,
       volume = {712},
       number = {2},
        pages = {1129-1136},
          doi = {10.1088/0004-637X/712/2/1129},
archivePrefix = {arXiv},
       eprint = {1002.4160},
 primaryClass = {astro-ph.CO},
       adsurl = {https://ui.adsabs.harvard.edu/abs/2010ApJ...712.1129M}
}

@ARTICLE{U2022a,
       author = {{U}, Vivian and {Barth}, Aaron J. and {Vogler}, H. Alexander and {Guo}, Hengxiao and {Treu}, Tommaso and {Bennert}, Vardha N. and {Canalizo}, Gabriela and {Filippenko}, Alexei V. and {Gates}, Elinor and {Hamann}, Frederick and {Joner}, Michael D. and {Malkan}, Matthew A. and {Pancoast}, Anna and {Williams}, Peter R. and {Woo}, Jong-Hak and {Abolfathi}, Bela and {Abramson}, L.~E. and {Armen}, Stephen F. and {Bae}, Hyun-Jin and {Bohn}, Thomas and {Boizelle}, Benjamin D. and {Bostroem}, Azalee and {Brandel}, Andrew and {Brink}, Thomas G. and {Channa}, Sanyum and {Cooper}, M.~C. and {Cosens}, Maren and {Donohue}, Edward and {Fillingham}, Sean P. and {Gonz{\'a}lez-Buitrago}, Diego and {Halevi}, Goni and {Halle}, Andrew and {Hood}, Carol E. and {Horne}, Keith and {Horst}, J. Chuck and {de Kouchkovsky}, Maxime and {Kuhn}, Benjamin and {Kumar}, Sahana and {Leonard}, Douglas C. and {Loveland}, Donald and {Manzano-King}, Christina and {McHardy}, Ian and {Michel}, Ra{\'u}l and {Olaes}, Melanie Kae B. and {Park}, Daeseong and {Park}, Songyoun and {Pei}, Liuyi and {Ross}, Timothy W. and {Runco}, Jordan N. and {Samuel}, Jenna and {S{\'a}nchez}, Javier and {Scott}, Bryan and {Sexton}, Remington O. and {Shin}, Jaejin and {Shivvers}, Isaac and {Spencer}, Chance L. and {Stahl}, Benjamin E. and {Stegman}, Samantha and {Stomberg}, Isak and {Valenti}, Stefano and {Villafa{\~n}a}, L. and {Walsh}, Jonelle L. and {Yuk}, Heechan and {Zheng}, WeiKang},
        title = "{The Lick AGN Monitoring Project 2016: Velocity-resolved H{\ensuremath{\beta}} Lags in Luminous Seyfert Galaxies}",
      journal = {\apj},
         year = 2022,
        month = jan,
       volume = {925},
       number = {1},
          eid = {52},
        pages = {52},
          doi = {10.3847/1538-4357/ac3d26},
archivePrefix = {arXiv},
       eprint = {2111.14849},
 primaryClass = {astro-ph.GA},
       adsurl = {https://ui.adsabs.harvard.edu/abs/2022ApJ...925...52U}
}

@ARTICLE{Schwarzschild1979,
       author = {{Schwarzschild}, M.},
        title = "{A numerical model for a triaxial stellar system in dynamical equilibrium.}",
      journal = {\apj},
         year = 1979,
        month = aug,
       volume = {232},
        pages = {236-247},
          doi = {10.1086/157282},
       adsurl = {https://ui.adsabs.harvard.edu/abs/1979ApJ...232..236S}
}

@ARTICLE{Wright2006,
       author = {{Wright}, E.~L.},
        title = "{A Cosmology Calculator for the World Wide Web}",
      journal = {\pasp},
         year = 2006,
        month = dec,
       volume = {118},
       number = {850},
        pages = {1711-1715},
          doi = {10.1086/510102},
archivePrefix = {arXiv},
       eprint = {astro-ph/0609593},
 primaryClass = {astro-ph},
       adsurl = {https://ui.adsabs.harvard.edu/abs/2006PASP..118.1711W}
}

@ARTICLE{Solomon1972,
       author = {{Solomon}, P.~M. and {Scoville}, N.~Z. and {Penzias}, A.~A. and {Wilson}, R.~W. and {Jefferts}, K.~B.},
        title = "{Molecular Clouds in the Galactic Center Region: Carbon Monoxide Observations at 2.6 Millimeters}",
      journal = {\apj},
         year = 1972,
        month = nov,
       volume = {178},
        pages = {125-130},
          doi = {10.1086/151772},
       adsurl = {https://ui.adsabs.harvard.edu/abs/1972ApJ...178..125S}
}

@ARTICLE{Wilson1974,
       author = {{Wilson}, W.~J. and {Schwartz}, P.~R. and {Epstein}, E.~E. and {Johnson}, W.~A. and {Etcheverry}, R.~D. and {Mori}, T.~T. and {Berry}, G.~G. and {Dyson}, H.~B.},
        title = "{Observations of Galactic Carbon Monoxide Emission at 2.6 Millimeters}",
      journal = {\apj},
         year = 1974,
        month = jul,
       volume = {191},
        pages = {357-374},
          doi = {10.1086/152974},
       adsurl = {https://ui.adsabs.harvard.edu/abs/1974ApJ...191..357W}
}

@ARTICLE{ScovilleSolomon1975,
       author = {{Scoville}, N.~Z. and {Solomon}, P.~M.},
        title = "{Molecular clouds in the Galaxy.}",
      journal = {\apjl},
         year = 1975,
        month = jul,
       volume = {199},
        pages = {L105-L109},
          doi = {10.1086/181859},
       adsurl = {https://ui.adsabs.harvard.edu/abs/1975ApJ...199L.105S}
}

@ARTICLE{Burton1975,
       author = {{Burton}, W.~B. and {Gordon}, M.~A. and {Bania}, T.~M. and {Lockman}, F.~J.},
        title = "{The overall distribution of carbon monoxide in the plane of the Galaxy.}",
      journal = {\apj},
         year = 1975,
        month = nov,
       volume = {202},
        pages = {30-49},
          doi = {10.1086/153950},
       adsurl = {https://ui.adsabs.harvard.edu/abs/1975ApJ...202...30B}
}

@ARTICLE{Magorrian1998,
       author = {{Magorrian}, John and {Tremaine}, Scott and {Richstone}, Douglas and {Bender}, Ralf and {Bower}, Gary and {Dressler}, Alan and {Faber}, S.~M. and {Gebhardt}, Karl and {Green}, Richard and {Grillmair}, Carl and {Kormendy}, John and {Lauer}, Tod},
        title = "{The Demography of Massive Dark Objects in Galaxy Centers}",
      journal = {\aj},
         year = 1998,
        month = jun,
       volume = {115},
       number = {6},
        pages = {2285-2305},
          doi = {10.1086/300353},
archivePrefix = {arXiv},
       eprint = {astro-ph/9708072},
 primaryClass = {astro-ph},
       adsurl = {https://ui.adsabs.harvard.edu/abs/1998AJ....115.2285M}
}

@ARTICLE{Khan2016,
       author = {{Khan}, Fazeel Mahmood and {Fiacconi}, Davide and {Mayer}, Lucio and {Berczik}, Peter and {Just}, Andreas},
        title = "{Swift Coalescence of Supermassive Black Holes in Cosmological Mergers of Massive Galaxies}",
      journal = {\apj},
         year = 2016,
        month = sep,
       volume = {828},
       number = {2},
          eid = {73},
        pages = {73},
          doi = {10.3847/0004-637X/828/2/73},
archivePrefix = {arXiv},
       eprint = {1604.00015},
 primaryClass = {astro-ph.GA},
       adsurl = {https://ui.adsabs.harvard.edu/abs/2016ApJ...828...73K}
}

@ARTICLE{SandersMirabel1996,
       author = {{Sanders}, D.~B. and {Mirabel}, I.~F.},
        title = "{Luminous Infrared Galaxies}",
      journal = {\araa},
         year = 1996,
        month = jan,
       volume = {34},
        pages = {749},
          doi = {10.1146/annurev.astro.34.1.749},
       adsurl = {https://ui.adsabs.harvard.edu/abs/1996ARA&A..34..749S}
}

@ARTICLE{Montoya2023,
       author = {{Montoya Arroyave}, I. and {Cicone}, C. and {Makroleivaditi}, E. and {Weiss}, A. and {Lundgren}, A. and {Severgnini}, P. and {De Breuck}, C. and {Baumschlager}, B. and {Schimek}, A. and {Shen}, S. and {Aravena}, M.},
        title = "{A sensitive APEX and ALMA CO(1-0), CO(2-1), CO(3-2), and [CI](1-0) spectral survey of 40 local (ultra-)luminous infrared galaxies}",
      journal = {\aap},
         year = 2023,
        month = may,
       volume = {673},
          eid = {A13},
        pages = {A13},
          doi = {10.1051/0004-6361/202245046},
archivePrefix = {arXiv},
       eprint = {2302.06629},
 primaryClass = {astro-ph.GA},
       adsurl = {https://ui.adsabs.harvard.edu/abs/2023A&A...673A..13M}
}

@ARTICLE{Herrero2019,
       author = {{Herrero-Illana}, R. and {Privon}, G.~C. and {Evans}, A.~S. and {D{\'\i}az-Santos}, T. and {P{\'e}rez-Torres}, M. {\'A}. and {U}, V. and {Alberdi}, A. and {Iwasawa}, K. and {Armus}, L. and {Aalto}, S. and {Mazzarella}, J. and {Chu}, J. and {Sanders}, D.~B. and {Barcos-Mu{\~n}oz}, L. and {Charmandaris}, V. and {Linden}, S.~T. and {Yoon}, I. and {Frayer}, D.~T. and {Inami}, H. and {Kim}, D. -C. and {Borish}, H.~J. and {Conway}, J. and {Murphy}, E.~J. and {Song}, Y. and {Stierwalt}, S. and {Surace}, J.},
        title = "{Molecular gas and dust properties of galaxies from the Great Observatories All-sky LIRG Survey}",
      journal = {\aap},
         year = 2019,
        month = aug,
       volume = {628},
          eid = {A71},
        pages = {A71},
          doi = {10.1051/0004-6361/201834088},
archivePrefix = {arXiv},
       eprint = {1907.03854},
 primaryClass = {astro-ph.GA},
       adsurl = {https://ui.adsabs.harvard.edu/abs/2019A&A...628A..71H}
}

@ARTICLE{CASA2022,
       author = {{CASA Team} and {Bean}, Ben and {Bhatnagar}, Sanjay and {Castro}, Sandra and {Donovan Meyer}, Jennifer and {Emonts}, Bjorn and {Garcia}, Enrique and {Garwood}, Robert and {Golap}, Kumar and {Gonzalez Villalba}, Justo and {Harris}, Pamela and {Hayashi}, Yohei and {Hoskins}, Josh and {Hsieh}, Mingyu and {Jagannathan}, Preshanth and {Kawasaki}, Wataru and {Keimpema}, Aard and {Kettenis}, Mark and {Lopez}, Jorge and {Marvil}, Joshua and {Masters}, Joseph and {McNichols}, Andrew and {Mehringer}, David and {Miel}, Renaud and {Moellenbrock}, George and {Montesino}, Federico and {Nakazato}, Takeshi and {Ott}, Juergen and {Petry}, Dirk and {Pokorny}, Martin and {Raba}, Ryan and {Rau}, Urvashi and {Schiebel}, Darrell and {Schweighart}, Neal and {Sekhar}, Srikrishna and {Shimada}, Kazuhiko and {Small}, Des and {Steeb}, Jan-Willem and {Sugimoto}, Kanako and {Suoranta}, Ville and {Tsutsumi}, Takahiro and {van Bemmel}, Ilse M. and {Verkouter}, Marjolein and {Wells}, Akeem and {Xiong}, Wei and {Szomoru}, Arpad and {Griffith}, Morgan and {Glendenning}, Brian and {Kern}, Jeff},
        title = "{CASA, the Common Astronomy Software Applications for Radio Astronomy}",
      journal = {\pasp},
         year = 2022,
        month = nov,
       volume = {134},
       number = {1041},
          eid = {114501},
        pages = {114501},
          doi = {10.1088/1538-3873/ac9642},
archivePrefix = {arXiv},
       eprint = {2210.02276},
 primaryClass = {astro-ph.IM},
       adsurl = {https://ui.adsabs.harvard.edu/abs/2022PASP..134k4501C}
}

@ARTICLE{Angles2017,
       author = {{Angl{\'e}s-Alc{\'a}zar}, Daniel and {Faucher-Gigu{\`e}re}, Claude-Andr{\'e} and {Quataert}, Eliot and {Hopkins}, Philip F. and {Feldmann}, Robert and {Torrey}, Paul and {Wetzel}, Andrew and {Kere{\v{s}}}, Du{\v{s}}an},
        title = "{Black holes on FIRE: stellar feedback limits early feeding of galactic nuclei}",
      journal = {\mnras},
         year = 2017,
        month = nov,
       volume = {472},
       number = {1},
        pages = {L109-L114},
          doi = {10.1093/mnrasl/slx161},
archivePrefix = {arXiv},
       eprint = {1707.03832},
 primaryClass = {astro-ph.GA},
       adsurl = {https://ui.adsabs.harvard.edu/abs/2017MNRAS.472L.109A}
}

@ARTICLE{Bennert2011,
       author = {{Bennert}, Vardha N. and {Auger}, Matthew W. and {Treu}, Tommaso and {Woo}, Jong-Hak and {Malkan}, Matthew A.},
        title = "{The Relation between Black Hole Mass and Host Spheroid Stellar Mass Out to z \raisebox{-0.5ex}\textasciitilde 2}",
      journal = {\apj},
         year = 2011,
        month = dec,
       volume = {742},
       number = {2},
          eid = {107},
        pages = {107},
          doi = {10.1088/0004-637X/742/2/107},
archivePrefix = {arXiv},
       eprint = {1102.1975},
 primaryClass = {astro-ph.CO},
       adsurl = {https://ui.adsabs.harvard.edu/abs/2011ApJ...742..107B}
}

@ARTICLE{Cisternas2011a,
       author = {{Cisternas}, Mauricio and {Jahnke}, Knud and {Bongiorno}, Angela and {Inskip}, Katherine J. and {Impey}, Chris D. and {Koekemoer}, Anton M. and {Merloni}, Andrea and {Salvato}, Mara and {Trump}, Jonathan R.},
        title = "{Secular Evolution and a Non-evolving Black-hole-to-galaxy Mass Ratio in the Last 7 Gyr}",
      journal = {\apjl},
         year = 2011,
        month = nov,
       volume = {741},
       number = {1},
          eid = {L11},
        pages = {L11},
          doi = {10.1088/2041-8205/741/1/L11},
archivePrefix = {arXiv},
       eprint = {1109.4633},
 primaryClass = {astro-ph.CO},
       adsurl = {https://ui.adsabs.harvard.edu/abs/2011ApJ...741L..11C}
}

@ARTICLE{Heckman2014,
       author = {{Heckman}, Timothy M. and {Best}, Philip N.},
        title = "{The Coevolution of Galaxies and Supermassive Black Holes: Insights from Surveys of the Contemporary Universe}",
      journal = {\araa},
         year = 2014,
        month = aug,
       volume = {52},
        pages = {589-660},
          doi = {10.1146/annurev-astro-081913-035722},
archivePrefix = {arXiv},
       eprint = {1403.4620},
 primaryClass = {astro-ph.GA},
       adsurl = {https://ui.adsabs.harvard.edu/abs/2014ARA&A..52..589H}
}

@article{astropy:2013,
Adsurl = {http://adsabs.harvard.edu/abs/2013A%26A...558A..33A},
Archiveprefix = {arXiv},
Author = {{Astropy Collaboration} and {Robitaille}, T.~P. and {Tollerud}, E.~J. and {Greenfield}, P. and {Droettboom}, M. and {Bray}, E. and {Aldcroft}, T. and {Davis}, M. and {Ginsburg}, A. and {Price-Whelan}, A.~M. and {Kerzendorf}, W.~E. and {Conley}, A. and {Crighton}, N. and {Barbary}, K. and {Muna}, D. and {Ferguson}, H. and {Grollier}, F. and {Parikh}, M.~M. and {Nair}, P.~H. and {Unther}, H.~M. and {Deil}, C. and {Woillez}, J. and {Conseil}, S. and {Kramer}, R. and {Turner}, J.~E.~H. and {Singer}, L. and {Fox}, R. and {Weaver}, B.~A. and {Zabalza}, V. and {Edwards}, Z.~I. and {Azalee Bostroem}, K. and {Burke}, D.~J. and {Casey}, A.~R. and {Crawford}, S.~M. and {Dencheva}, N. and {Ely}, J. and {Jenness}, T. and {Labrie}, K. and {Lim}, P.~L. and {Pierfederici}, F. and {Pontzen}, A. and {Ptak}, A. and {Refsdal}, B. and {Servillat}, M. and {Streicher}, O.},
Doi = {10.1051/0004-6361/201322068},
Eid = {A33},
Eprint = {1307.6212},
Journal = {\aap},
Month = oct,
Pages = {A33},
Primaryclass = {astro-ph.IM},
Title = {{Astropy: A community Python package for astronomy}},
Volume = 558,
Year = 2013}

@ARTICLE{astropy:2018,
       author = {{Astropy Collaboration} and {Price-Whelan}, A.~M. and
         {Sip{\H{o}}cz}, B.~M. and {G{\"u}nther}, H.~M. and {Lim}, P.~L. and
         {Crawford}, S.~M. and {Conseil}, S. and {Shupe}, D.~L. and
         {Craig}, M.~W. and {Dencheva}, N. and {Ginsburg}, A. and {Vand
        erPlas}, J.~T. and {Bradley}, L.~D. and {P{\'e}rez-Su{\'a}rez}, D. and
         {de Val-Borro}, M. and {Aldcroft}, T.~L. and {Cruz}, K.~L. and
         {Robitaille}, T.~P. and {Tollerud}, E.~J. and {Ardelean}, C. and
         {Babej}, T. and {Bach}, Y.~P. and {Bachetti}, M. and {Bakanov}, A.~V. and
         {Bamford}, S.~P. and {Barentsen}, G. and {Barmby}, P. and
         {Baumbach}, A. and {Berry}, K.~L. and {Biscani}, F. and {Boquien}, M. and
         {Bostroem}, K.~A. and {Bouma}, L.~G. and {Brammer}, G.~B. and
         {Bray}, E.~M. and {Breytenbach}, H. and {Buddelmeijer}, H. and
         {Burke}, D.~J. and {Calderone}, G. and {Cano Rodr{\'\i}guez}, J.~L. and
         {Cara}, M. and {Cardoso}, J.~V.~M. and {Cheedella}, S. and {Copin}, Y. and
         {Corrales}, L. and {Crichton}, D. and {D'Avella}, D. and {Deil}, C. and
         {Depagne}, {\'E}. and {Dietrich}, J.~P. and {Donath}, A. and
         {Droettboom}, M. and {Earl}, N. and {Erben}, T. and {Fabbro}, S. and
         {Ferreira}, L.~A. and {Finethy}, T. and {Fox}, R.~T. and
         {Garrison}, L.~H. and {Gibbons}, S.~L.~J. and {Goldstein}, D.~A. and
         {Gommers}, R. and {Greco}, J.~P. and {Greenfield}, P. and
         {Groener}, A.~M. and {Grollier}, F. and {Hagen}, A. and {Hirst}, P. and
         {Homeier}, D. and {Horton}, A.~J. and {Hosseinzadeh}, G. and {Hu}, L. and
         {Hunkeler}, J.~S. and {Ivezi{\'c}}, {\v{Z}}. and {Jain}, A. and
         {Jenness}, T. and {Kanarek}, G. and {Kendrew}, S. and {Kern}, N.~S. and
         {Kerzendorf}, W.~E. and {Khvalko}, A. and {King}, J. and {Kirkby}, D. and
         {Kulkarni}, A.~M. and {Kumar}, A. and {Lee}, A. and {Lenz}, D. and
         {Littlefair}, S.~P. and {Ma}, Z. and {Macleod}, D.~M. and
         {Mastropietro}, M. and {McCully}, C. and {Montagnac}, S. and
         {Morris}, B.~M. and {Mueller}, M. and {Mumford}, S.~J. and {Muna}, D. and
         {Murphy}, N.~A. and {Nelson}, S. and {Nguyen}, G.~H. and
         {Ninan}, J.~P. and {N{\"o}the}, M. and {Ogaz}, S. and {Oh}, S. and
         {Parejko}, J.~K. and {Parley}, N. and {Pascual}, S. and {Patil}, R. and
         {Patil}, A.~A. and {Plunkett}, A.~L. and {Prochaska}, J.~X. and
         {Rastogi}, T. and {Reddy Janga}, V. and {Sabater}, J. and
         {Sakurikar}, P. and {Seifert}, M. and {Sherbert}, L.~E. and
         {Sherwood-Taylor}, H. and {Shih}, A.~Y. and {Sick}, J. and
         {Silbiger}, M.~T. and {Singanamalla}, S. and {Singer}, L.~P. and
         {Sladen}, P.~H. and {Sooley}, K.~A. and {Sornarajah}, S. and
         {Streicher}, O. and {Teuben}, P. and {Thomas}, S.~W. and
         {Tremblay}, G.~R. and {Turner}, J.~E.~H. and {Terr{\'o}n}, V. and
         {van Kerkwijk}, M.~H. and {de la Vega}, A. and {Watkins}, L.~L. and
         {Weaver}, B.~A. and {Whitmore}, J.~B. and {Woillez}, J. and
         {Zabalza}, V. and {Astropy Contributors}},
        title = "{The Astropy Project: Building an Open-science Project and Status of the v2.0 Core Package}",
      journal = {\aj},
         year = 2018,
        month = sep,
       volume = {156},
       number = {3},
          eid = {123},
        pages = {123},
          doi = {10.3847/1538-3881/aabc4f},
archivePrefix = {arXiv},
       eprint = {1801.02634},
 primaryClass = {astro-ph.IM},
       adsurl = {https://ui.adsabs.harvard.edu/abs/2018AJ....156..123A}
}

@ARTICLE{Bennert2021,
       author = {{Bennert}, Vardha N. and {Treu}, Tommaso and {Ding}, Xuheng and {Stomberg}, Isak and {Birrer}, Simon and {Snyder}, Tomas and {Malkan}, Matthew A. and {Stephens}, Andrew W. and {Auger}, Matthew W.},
        title = "{A Local Baseline of the Black Hole Mass Scaling Relations for Active Galaxies. IV. Correlations Between M$_{BH}$ and Host Galaxy {\ensuremath{\sigma}}, Stellar Mass, and Luminosity}",
      journal = {\apj},
         year = 2021,
        month = nov,
       volume = {921},
       number = {1},
          eid = {36},
        pages = {36},
          doi = {10.3847/1538-4357/ac151a},
archivePrefix = {arXiv},
       eprint = {2101.10355},
 primaryClass = {astro-ph.GA},
       adsurl = {https://ui.adsabs.harvard.edu/abs/2021ApJ...921...36B}
}

@ARTICLE{Wang2011,
       author = {{Wang}, Junzhi and {Zhang}, Zhiyu and {Shi}, Yong},
        title = "{CS (5-4) survey towards nearby infrared bright galaxies}",
      journal = {\mnras},
         year = 2011,
        month = sep,
       volume = {416},
       number = {1},
        pages = {L21-L25},
          doi = {10.1111/j.1745-3933.2011.01090.x},
archivePrefix = {arXiv},
       eprint = {1106.2873},
 primaryClass = {astro-ph.CO},
       adsurl = {https://ui.adsabs.harvard.edu/abs/2011MNRAS.416L..21W}
}

@software{CARTA,
       author = {{Comrie}, Angus and {Wang}, Kuo-Song and {Hsu}, Shou-Chieh and {Moraghan}, Anthony and {Harris}, Pamela and {Pang}, Qi and {Pi{\'n}ska}, Adrianna and {Chiang}, Cheng-Chin and {Chang}, Tien-Hao and {Hwang}, Yu-Hsuan and {Jan}, Hengtai and {Lin}, Ming-Yi and {Simmonds}, Rob},
        title = "{CARTA: The Cube Analysis and Rendering Tool for Astronomy}",
         year = 2021,
        month = jun,
          eid = {10.5281/zenodo.4905459},
          doi = {10.5281/zenodo.4905459},
      version = {2.0.0},
    publisher = {Zenodo},
       adsurl = {https://ui.adsabs.harvard.edu/abs/2021zndo...4905459C}
}

@ARTICLE{Teng2022,
       author = {{Teng}, Yu-Hsuan and {Sandstrom}, Karin M. and {Sun}, Jiayi and {Leroy}, Adam K. and {Johnson}, L. Clifton and {Bolatto}, Alberto D. and {Kruijssen}, J.~M. Diederik and {Schruba}, Andreas and {Usero}, Antonio and {Barnes}, Ashley T. and {Bigiel}, Frank and {Blanc}, Guillermo A. and {Groves}, Brent and {Israel}, Frank P. and {Liu}, Daizhong and {Rosolowsky}, Erik and {Schinnerer}, Eva and {Smith}, J.~D. and {Walter}, Fabian},
        title = "{Molecular Gas Properties and CO-to-H$_{2}$ Conversion Factors in the Central Kiloparsec of NGC 3351}",
      journal = {\apj},
         year = 2022,
        month = jan,
       volume = {925},
       number = {1},
          eid = {72},
        pages = {72},
          doi = {10.3847/1538-4357/ac382f},
archivePrefix = {arXiv},
       eprint = {2111.05844},
 primaryClass = {astro-ph.GA},
       adsurl = {https://ui.adsabs.harvard.edu/abs/2022ApJ...925...72T}
}

\begin{appendix}

\section{Kinematic Modeling}
\label{app:kinematic_modeling}
This section of the Appendix describes additional detail about the $^{3\mathrm{D}}$Barolo tilted-ring modeling used in this work to calculate \menc\ from the cold CO gas. $^{3\mathrm{D}}$Barolo is a tilted-ring modeling algorithm with a task called \verb|3DFIT|. This algorithm takes in data with kinematic information, and derives rotational velocities for tilted rings spaced at intervals specified by the user. For our purposes with the high resolution CO(2-1) cubes, we use \verb|3DFIT| in two-stage mode. In two-stage mode \verb|3DFIT| regularizes the input parameters, then enters a second fitting phase (\citealt{3DBarolo}). 

We tested different methods of building the \verb|3DFIT| mask (used by the program to determine where there is genuine emission), using both SEARCH and SMOOTH\&SEARCH, and we tested different input parameters (e.g. PA, inclination, redshift, and systemic velocity). In testing, as is suggested by the authors of $^{3\mathrm{D}}$Barolo, we judge model goodness by examining the line intensity maps and PV plots (as in Figure~\ref{fig:PVdia}).
Models and residuals produced in this way are shown in Figures~\ref{bbarolo3Zw} and~\ref{bbaroloIR01364}, and output tilted-ring parameters can be found in Table~\ref{table:3DBarolooutputs}. We use the resulting rotational velocities in Section~\ref{sec:PVdiagrams} to calculate \menc.

\begin{table}[ht]
    \centering
    \hspace{-2.5cm}
    \scalebox{0.80}{
    \begin{tabular}{c|c|c|c|c}
         Galaxy & R (arcsec) & v$_{\mathrm{rot}}$ (km s$^{-1}$) & $\sigma_\mathrm{v}$ (km s$^{-1}$) & P.A. (deg)\\
         \hline
         \zw & 0.010 & 117.299 $\substack{+15.688 \\ -16.559}$ & 12.962 $\substack{+ 13.157 \\ -13.157}$ & 19.765   \\
          \hline
           & 0.030 & 116.632 $\substack{+24.541 \\ -23.268}$  & 70.446 $\substack{+13.758 \\ -13.758}$ & 19.765 \\
          \hline
          & 0.050 & 127.537 $\substack{+17.974 \\ -15.995}$ & 71.341 $\substack{+13.321 \\ -11.454}$ & 19.765   \\
          \hline
          & 0.070 & 138.887 \textbf{$\substack{+14.576 \\ -13.268}$ } & 64.466 $\substack{+11.887 \\ -9.568}$ & 19.765 \\
          \hline
          & 0.090 & 155.569 $\substack{+13.731 \\ -15.584}$ & 66.102 $\substack{+11.342 \\ -12.737}$ & 19.765  \\
          \hline
          & 0.110 & 174.523 $\substack{+20.043 \\ -21.25}$ & 67.762 $\substack{+14.458 \\ -14.458}$ & 19.765  \\
          \hline
          & 0.130 & 184.272 $\substack{+19.313 \\ -19.348}$ & 58.828 $\substack{+12.395 \\ -12.395}$ & 19.765 \\
          \hline
          & 0.150 & 195.405 $\substack{+21.794 \\ -22.034}$ & 60.046 $\substack{+12.700 \\ -12.700}$ & 19.765 \\
          \hline
          & 0.170 & 206.447 $\substack{+24.747 \\ -26.912}$ & 56.457 $\substack{+12.373 \\ -13.456}$ & 19.765  \\
          \hline
          & 0.190 & 210.448 $\substack{+28.939 \\ -26.236}$ & 44.030 $\substack{+14.469 \\ -13.118}$ & 19.765  \\
          \hline
         \iras& 0.013 & 158.335 $\substack{+21.370 \\ -17.959}$  & 30.945 $\substack{+11.350 \\ -11.103}$ & 240.958 \\
          \hline
          & 0.039 & 150.972 $\substack{+18.810 \\ -22.804}$ & 65.695 $\substack{+14.215 \\ -14.215}$ & 236.849  \\
          \hline
          & 0.065 & 162.998 $\substack{+13.971 \\ -14.571}$ & 61.799 $\substack{+9.680 \\ -9.638}$ & 227.546 \\
          \hline
          & 0.091 & 173.123 $\substack{+13.000 \\ -12.731}$ & 62.743 $\substack{+8.393 \\ -8.159}$ & 223.659  \\
          \hline
          & 0.117 & 182.589 $\substack{+11.929 \\ -13.586}$ & 63.689 $\substack{+8.303 \\ -9.952}$ & 224.687 \\
          \hline
          & 0.143 & 180.143 $\substack{+15.618 \\ -15.427}$ & 64.256 $\substack{+10.198 \\ -10.775}$ & 227.474  \\
          \hline
          & 0.169 & 168.557 $\substack{+18.185 \\ -18.222}$  & 61.354 $\substack{+10.996 \\ -11.841}$ & 229.847 \\
          \hline
          & 0.195 & 152.269 $\substack{+22.261 \\ -20.124}$ & 57.424 $\substack{+12.634 \\ -13.882}$ & 231.254 \\
          \hline
          & 0.221 & 133.305 $\substack{+23.109 \\ -22.004}$  & 53.216 $\substack{+11.214 \\ -11.214}$ & 232.371  \\
          \hline
          & 0.247 & 137.034 $\substack{+23.665 \\ -26.155}$ & 45.981 $\substack{+12.003 \\ -12.003}$ & 235.257 \\
    \end{tabular}}
    \caption{$^{3\mathrm{D}}$Barolo phase two best-fit tilted-ring model parameters for \zw\ and \iras. The velocity dispersion modeling (column 4) may be somewhat elevated due to contamination of the CO line profiles by an outflowing component. The second and third columns, tilted-ring radius and rotational velocity, are used to compute \menc\ as found in Section~\ref{sec:PVdiagrams}. The inclination angles for both galaxies were constant as a function of radius: 75 and 61 degrees for \zw\ and \iras\, respectively. Position angle (column 5), while constant for \zw, was fit as a free variable for both galaxies.}
    \label{table:3DBarolooutputs}
\end{table}

\begin{figure*}[ht]
\centering
\includegraphics[width=0.75\textwidth]{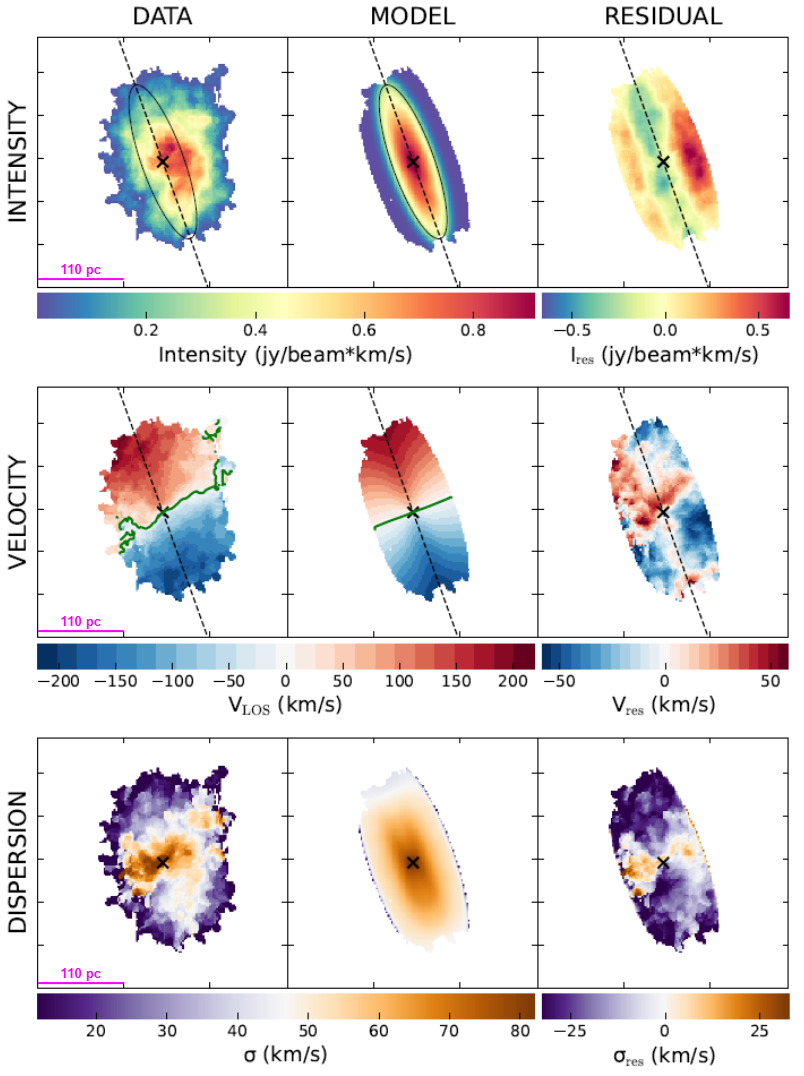}
\caption{\zw\ ring models produced by $^{3\mathrm{D}}$Barolo, used to calculate \menc\ in Section~\ref{sec:PVdiagrams}. Input CO(2-1) images are the same as presented in Figure~\ref{fig:COims} (33 $\times$ 28 mas beam size), and input parameters include PA, inclination, redshift, and systemic velocity. Ring model output parameters can be found in Table~\ref{table:3DBarolooutputs}. Green lines in the first two velocity panels indicate 0 km s$^{-1}$ with respect to the galaxy's systemic velocity, which indicate potential disk-warping. The higher velocity dispersion residuals of the \zw\ disk along its minor axis likely reflect broadening of the CO line profile by the molecular outflow.}
\label{bbarolo3Zw}
\end{figure*}
\newpage
\begin{figure*}[ht]
\centering
\includegraphics[width=0.75\textwidth]{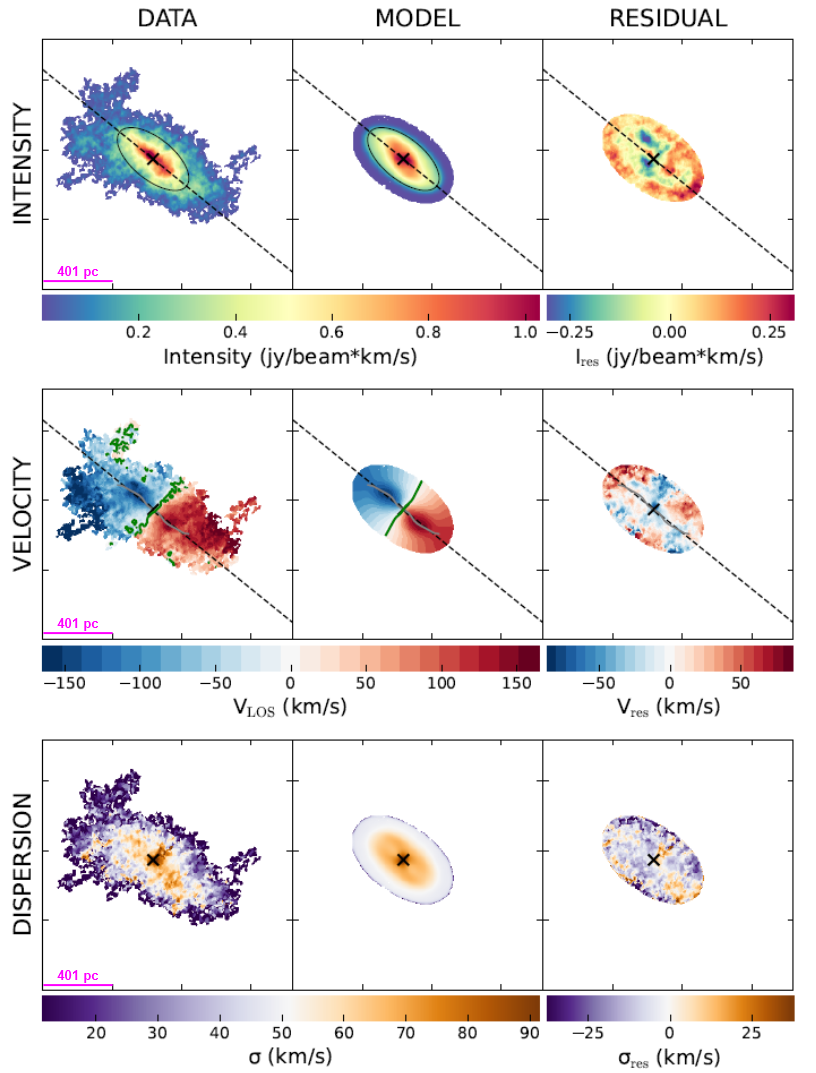}
\caption{\iras\ ring models produced by $^{3\mathrm{D}}$Barolo, used to calculate \menc\ in Section~\ref{sec:PVdiagrams}. Input CO(2-1) images are the same as presented in Figure~\ref{fig:COims} (42 $\times$ 39 mas beam size), and input parameters include PA, inclination, redshift, and systemic velocity. Ring model output parameters can be found in Table~\ref{table:3DBarolooutputs}. Green lines in the first two velocity panels indicate 0 km s$^{-1}$ with respect to the galaxy's systemic velocity. \iras\ is particularly well-defined by a rotating disk, although residuals show some evidence of outflows along the minor axis in velocity space.}
\label{bbaroloIR01364}
\end{figure*}

\end{appendix}

\end{document}